\documentclass[reprint,amsmath,amssymb,aps,prd,superscriptaddress,nofootinbib,floatfix]{revtex4-2}

\usepackage{graphicx}
\usepackage{bm}
\usepackage{amsmath}
\usepackage{amssymb}
\usepackage{booktabs}
\usepackage{xcolor}
\usepackage{tikz}
\usepackage[compat=1.1.0]{tikz-feynman}  
\usepackage[colorlinks=true,allcolors=blue]{hyperref}

\newcommand{\CHF}{\ensuremath{\mathrm{C_{2}H_{2}F_{4}}}}   
\newcommand{\InDEx}{\mbox{InDEx}}
\newcommand{\mchi}{m_{\chi}}
\newcommand{\me}{m_{e}}
\newcommand{\mue}{\mu_{\chi e}}
\newcommand{\sigmae}{\overline{\sigma}_{e}}
\newcommand{\Fdm}{F_{\mathrm{DM}}}
\newcommand{\vmin}{v_{\mathrm{min}}}
\newcommand{\Ee}{E_{e}}
\newcommand{\aem}{\alpha}
\newcommand{\fion}{f_{\mathrm{ion}}}
\newcommand{\rhochi}{\rho_{\chi}}

\begin{document}

\title{Probing sub-GeV dark matter through dark matter--electron
       scattering\\ in the superheated \CHF{} target of the \InDEx{}
       experiment}

\author{Utkarsh Patel}
\email{utkarsh.patel@saha.ac.in}
\affiliation{Nuclear Atomic \& Particle Physics Division,\\ Saha Institute of Nuclear Physics, 1/AF Bidhannagar, Kolkata 700064, India}

\author{Astitva Kathait}
\email{astitva.kathait@saha.ac.in}
\affiliation{Nuclear Atomic \& Particle Physics Division,\\ Saha Institute of Nuclear Physics, 1/AF Bidhannagar, Kolkata 700064, India}
\affiliation{Homi Bhabha National Institute, Training School Complex,\\ Anushakti Nagar, Mumbai 400094, India}

\author{Mala Das}
\email{mala.das@saha.ac.in}
\affiliation{Nuclear Atomic \& Particle Physics Division,\\ Saha Institute of Nuclear Physics, 1/AF Bidhannagar, Kolkata 700064, India}
\affiliation{Homi Bhabha National Institute, Training School Complex,\\ Anushakti Nagar, Mumbai 400094, India}

\date{\today}

\begin{abstract}
The Indian Dark Matter Search Experiment (\InDEx) operates superheated droplet detectors with tetrafluoroethane (\CHF) as the active target at the Jaduguda Underground Science Laboratory (JUSL), and has so far probed the dark matter (DM) through nuclear recoils at operating temperatures at which the target is insensitive to electron recoils. In this work, we consider six operating temperatures, $T=45$--$70~^{\circ}$C in steps of $5~^{\circ}$C; the corresponding Seitz thresholds range from $0.61~\mathrm{keV}$ down to $26.8~\mathrm{eV}$. The ionization response of the \CHF{} target is computed in the isolated-atom approximation from Roothaan--Hartree--Fock wavefunctions. The projected $90\%$ confidence level upper limits on the DM--electron scattering cross section are derived for a dark-photon-type mediator, in both the heavy- and light-mediator limits, for a benchmark exposure of $10^{3}~\mathrm{kg\, days}$ under the zero-background assumption. For the lowest threshold, $E_{\rm th}=26.8$~eV at $T=70~^{\circ}$C, the search probes DM masses down to $\sim12$~MeV, with the $90\%$ confidence level limit reaching $2.7\times10^{-41}~\mathrm{cm}^{2}$ near $\mchi\simeq100$~MeV for the contact interaction and $3.0\times10^{-37}~\mathrm{cm}^{2}$ near $\mchi\simeq83$~MeV for the long-range interaction. The electron channel thus extends the reach of \InDEx{} to sub-GeV DM masses, down to $\mathcal{O}(10\,\mathrm{MeV})$, providing a complementary probe of low-mass dark matter alongside its nuclear-recoil program and enhancing sensitivity to DM candidates coupling predominantly to leptons.
\end{abstract}

\maketitle

\section{Introduction}\label{sec:intro}
A substantial body of astrophysical and cosmological evidence indicates that about $85\%$ of the matter content of the Universe is non-luminous and non-baryonic~\cite{bertone2005,planck2018}. The canonical direct-detection strategy searches for the nuclear recoils induced by the elastic scattering of Galactic weakly interacting massive particles (WIMPs) off detector nuclei~\cite{goodman1985,drukier1986,lewin1996}. Multi-tonne liquid-xenon time-projection chambers currently lead this program, with the LZ~\cite{Mount:2017qzi}, XENONnT~\cite{XENON:2024wpa}, and PandaX-4T~\cite{PandaX-4T:2021bab} experiments constraining the spin-independent WIMP--nucleon cross section at the level of a few $\times10^{-48}~\mathrm{cm}^{2}$ for masses near $25$--$30~\mathrm{GeV}/c^{2}$~\cite{LZ:2024zvo,XENON:2023cxc,PandaX:2024qfu}. The absence of a signal in this canonical mass window has intensified the effort towards lighter, sub-GeV DM candidates.

For DM masses below a few $\mathrm{GeV}/c^{2}$, elastic scattering off nuclei yields increasingly small recoil energies, making conventional nuclear-recoil searches progressively less sensitive~\cite{Essig:2011nj}.~In contrast, scattering off bound electrons remains effective, since even a sub-GeV halo particle can transfer sufficient energy to ionize an atom or excite an electron across a semiconductor band gap.~\cite{essig2012,graham2012,essig2016semic}. DM--electron scattering searches with noble-liquid targets~\cite{essig2012xenon10,XENON:2019gfn,DarkSide:2018bpj,DarkSide:2022knj,PandaX-II:2021nsg,PandaX:2022xqx,PandaX:2025rrz} and solid-state targets~\cite{SENSEI:2020dpa,DAMIC:2019dcn,DAMIC-M:2023gxo,DAMIC-M:2025luv,EDELWEISS:2020fxc,SuperCDMS:2018mne,SuperCDMS:2024yiv,CDEX:2022kcd} now probe DM masses from below an $\mathrm{MeV}/c^{2}$ up to a few $\mathrm{GeV}/c^{2}$, supported by a mature theoretical description of the target response~\cite{essig2012,essig2017,roberts2016,roberts2019,catena2020,catena2021crystal,catena2023beyond,liang2024}. Next-generation projects aim to push this reach substantially further. Oscura, building on the SENSEI skipper-CCD technology, plans an active mass of ${\sim}10$~kg of high-resistivity silicon skipper-CCD sensors fabricated on commercial 200~mm wafers, and projects, for a $30~\mathrm{kg\, yr}$ exposure, $90\%$~C.L.\ sensitivities of $\sigmae\sim10^{-43}~\mathrm{cm^{2}}$ for DM masses down to a few $\mathrm{MeV}/c^{2}$~\cite{Oscura:2022vmi,cervantes2023oscura}. DarkSPHERE, a proposed next-generation $3$~m-diameter spherical proportional counter of the NEWS-G collaboration, to be electro-formed underground at the Boulby Underground Laboratory (2840~m.w.e.) and operated with a He:i-C$_{4}$H$_{10}$ gas mixture at $5$~bar, targets a single-ionization-electron threshold and projects, for a background-free $300$-day exposure, sensitivities of $\sigmae\sim10^{-42}~\mathrm{cm^{2}}$ (heavy mediator) and $\sim10^{-40}~\mathrm{cm^{2}}$ (light mediator) for DM masses down to $\sim30$ and $\sim20~\mathrm{MeV}/c^{2}$, respectively~\cite{NEWS-G:2023qwh}.

The Indian Dark Matter Search Experiment (\InDEx) aims to search the low-mass DM with superheated droplet detectors (superheated emulsion detectors) using tetrafluoroethane, \CHF{} (b.p.\ $-26.3~^{\circ}$C), as the active target, operated at the Jaduguda Underground Science Laboratory (JUSL) under 555~m of rock overburden (1604~m water equivalent)~\cite{seth2020,Kumar:2025ofs,das2025index}. In run-1 the detectors were operated at an average temperature of $25.4~^{\circ}$C, corresponding to a bubble-nucleation threshold of $5.87$~keV, with an exposure of $2.47~\mathrm{kg\,days}$~\cite{Kumar:2025ofs}, in run-2 the threshold was lowered to $1.95$~keV by operating at $T=(35.0\pm0.1)~^{\circ}$C, with an exposure of $7.2~\mathrm{kg\,days}$~\cite{das2025index}. Both runs exploit a defining feature of the superheated-liquid technique: below the measured onset of gamma-ray-induced nucleation at $T=(38.5\pm1.4)~^{\circ}$C~\cite{sahoo2019gamma}, the target is blind to electron recoils, so the dominant ambient gamma backgrounds cannot nucleate bubbles in the nuclear-recoil searches~\cite{seth2020,Kumar:2025ofs,das2025index}.

Read in reverse, the same feature defines a second physics program for the same detectors. At temperatures above $T\approx40~^{\circ}$C, electron recoils are capable of nucleating bubbles~\cite{sahoo2019gamma}, allowing a DM particle scattering off a bound electron in the \CHF{} target to generate a detectable signal. In this work we quantify the corresponding physics reach: we compute the ionization response of the \CHF{} target, evaluate the DM--electron scattering rates for a dark-photon-type mediator~\cite{Holdom:1985ag,essig2012} in both the heavy-mediator (contact) and light-mediator (long-range) limits, and derive projected $90\%$~C.L.\ exclusion limits at six operating temperatures, $T=45$--$70~^{\circ}$C, for a benchmark exposure of $10^{3}~\mathrm{kg\, days}$. To our knowledge, this constitutes the first characterization of a superheated-liquid (bubble-nucleation) target in the DM--electron channel, adding a refrigerant-based target -- chemically and operationally distinct from the noble-liquid and semiconductor targets employed so far, to the landscape of sub-GeV DM searches. The theoretical treatment follows the standard ionization formalism of Refs.~\cite{essig2012,essig2017} and its general effective-theory formulation in Ref.~\cite{catena2020}, to which we refer for all derivations; only the final formulae needed to define the analysis are quoted here.

The remainder of this paper is organized as follows. Section~\ref{sec:detector} describes the detectors and motivates the choice of the operating-temperature window. Section~\ref{sec:signal} summarizes the expected DM--electron signal and the limit-setting procedure. Section~\ref{sec:results} presents the computed ionization response, event rates, and exclusion limits, and compares them with the current direct-detection landscape. We summarize in Sec.~\ref{sec:summary}. The details of the electronic-structure input used in the analysis are collected in Appendix~\ref{app:estructure}.

\section{The \texorpdfstring{\NoCaseChange{\InDEx{}}}{InDEx} detectors and the electron-channel operating window}\label{sec:detector}

The \InDEx{} detectors consist of drops of liquid \CHF{}, held in a metastable superheated state at ambient pressure, dispersed in an inert gel matrix~\cite{seth2020, Kumar:2025ofs, das2025index}. According to the Seitz ``heat-spike'' model~\cite{Seitz:1958}, a particle depositing energy in a droplet triggers the liquid-to-vapor transition if the energy deposited within a critical track length exceeds the critical energy $E_{c}(T, P)$ set by the thermodynamics of the liquid. The expanding vapor bubble emits an acoustic pulse that is recorded by piezoelectric sensors~\cite{seth2020,Kumar:2025ofs}. The device is therefore a \emph{threshold} detector, with an effective deposited-energy threshold that is continuously tuned by the operating temperature $T$ (and pressure). Raising $T$ at fixed pressure lowers the threshold. The Seitz thresholds of \CHF{} at atmospheric pressure, computed as in Ref.~\cite{seth2020} from standard reference-fluid thermodynamic data, fall steeply with temperature, from $E_{\mathrm{th}}=0.61~\mathrm{keV}$ at $T=45~^{\circ}$C to $E_{\mathrm{th}}=26.8~\mathrm{eV}$ at $T=70~^{\circ}$C, the six operating points used in this work are collected in Table~\ref{tab:exclusion}.

Gamma-calibration data of the PICO C$_3$F$_8$ bubble chambers favor a data-driven, ionization-based nucleation model whose threshold scales differently with the thermodynamic conditions~\cite{PICO:2019rsv}. In the absence of an equivalent calibration of \CHF{} across this temperature range, we adopt the Seitz critical energy $E_c(T)$ as a reasonably effective deposited-energy threshold for the electron channel, with unit detection efficiency above threshold, pending in-situ gamma calibrations of the \InDEx{} modules.

The operating temperature range, $T=45$--$70~^{\circ}$C, is chosen based on the electron-recoil response of the \CHF{} target. Gamma-induced bubble nucleation in superheated \CHF{} emulsions, mediated by the secondary electrons produced by gamma-ray interactions, has been observed to onset at $T=(38.5\pm1.4)~^{\circ}$C at atmospheric pressure~\cite{sahoo2019gamma}. Below this temperature, the target is blind to electronic interactions~\cite{sahoo2019gamma,derrico2000photon}, the very property exploited by the nuclear-recoil runs, operated at $T\leq35~^{\circ}$C~\cite{Kumar:2025ofs,das2025index}. The electron channel thus opens only above $T\approx40~^{\circ}$C, and we start electron-channel data taking at $T=45~^{\circ}$C, comfortably above the measured onset and its uncertainty. The upper end of the window is instead set by backgrounds: the same mechanism makes the detector respond with steeply increasing efficiency to the ambient gamma and beta fields as the superheat is raised~\cite{derrico2000photon,sahoo2019gamma,pico2019electron}. We adopt $T=70~^{\circ}$C -- where the Seitz threshold of $26.8~\mathrm{eV}$ already extends the kinematic reach deep into the sub-GeV regime, down to $\mchi\simeq12~\mathrm{MeV}$ (Sec.~\ref{sec:res-limits}) -- as the highest operating point at which the electron-recoil background remains practically manageable with the shielding and rock overburden available at JUSL, combined with the exposure-driven background statistics and the event-selection and analysis techniques established in the \InDEx{} runs~\cite{Kumar:2025ofs,das2025index}. The projections below accordingly assume zero background (Sec.~\ref{sec:limits}) and are to be read as best-case sensitivities.

\section{Expected dark matter--electron signal}\label{sec:signal}

\subsection{Ionization rate}\label{sec:rate}
We describe the local DM population by the Standard Halo Model, a truncated Maxwell--Boltzmann velocity distribution boosted to the detector frame~\cite{lewin1996}, with $\rhochi=0.4~\mathrm{GeV\,cm^{-3}}$, most probable speed $v_{0}=220~\mathrm{km\,s^{-1}}$, escape speed $v_{\mathrm{esc}}=544~\mathrm{km\,s^{-1}}$, and Earth speed $v_{E}=244~\mathrm{km\,s^{-1}}$ -- the standard parameter choices of the DM--electron scattering literature~\cite{essig2012,essig2017,catena2020}, adopted here to allow a direct comparison with published electron-channel limits.

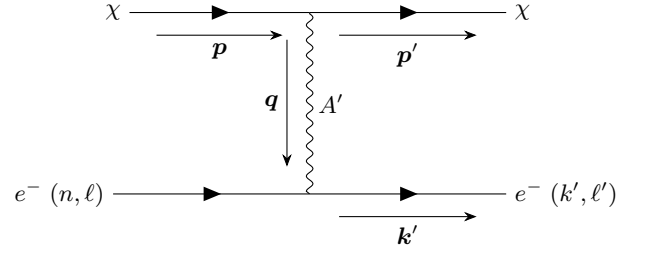
\begin{figure}[t]
  \centering
  \begin{tikzpicture}
    \begin{feynman}
      \vertex (a) {\(\chi\)};
      \vertex [right=2.6cm of a] (b);
      \vertex [right=2.6cm of b] (c) {\(\chi\)};
      \vertex [below=2.4cm of b] (d);
      \vertex [left=2.6cm of d] (e) {\(e^{-}\;(n,\ell)\)};
      \vertex [right=2.6cm of d] (f) {\(e^{-}\;(k',\ell')\)};
      \diagram* {
        (a) -- [fermion, momentum'=\(\bm{p}\)] (b)
            -- [fermion, momentum'=\(\bm{p}'\)] (c),
        (b) -- [boson, edge label=\(A'\), momentum'=\(\bm{q}\)] (d),
        (e) -- [fermion] (d)
            -- [fermion, momentum'=\(\bm{k}'\)] (f),
      };
    \end{feynman}
  \end{tikzpicture}
  \caption{Dark matter-induced ionization of a bound electron in the \CHF{} target. A halo DM particle $\chi$ of momentum $\bm{p}=\mchi\bm{v}$ transfers the momentum $\bm{q}=\bm{p}-\bm{p}'$ to an electron initially bound in an atomic orbital $(n,\ell)$ (which is therefore not in a momentum eigenstate and carries no momentum label), exciting it into a continuum state of asymptotic momentum $k'$ and angular momentum $\ell'$. For the dark-photon-type benchmark studied in this work, the exchanged mediator $A'$ is a vector boson kinetically mixed with the photon~\cite{Holdom:1985ag,essig2012}; the heavy- and light-mediator limits of its propagator generate the two DM form factors of Eq.~\eqref{eq:FDM}. Diagram drawn with \textsc{Ti\emph{k}Z-Feynman}~\cite{Ellis:2016jkw}.}
  \label{fig:feyn}
  \end{figure}

In DM-induced ionization, illustrated in Fig.~\ref{fig:feyn}, a halo DM particle $\chi$ of mass $\mchi$ and velocity $v$ lifts an electron from an orbital of binding energy $E_{B}$ into a continuum state of kinetic energy $\Ee$, depositing the energy $\Delta E=E_{B}+\Ee$. Because the initial electron is bound, the momentum transfer $q$ and $\Delta E$ are independent variables, and the minimum DM speed capable of transferring $(q,\Delta E)$ is~\cite{essig2012,catena2020}
\begin{equation}
\vmin(q,\Ee) \;=\; \frac{E_{B}+\Ee}{q} \;+\; \frac{q}{2\mchi}\,,
\label{eq:vmin}
\end{equation}
which enters the rate through the mean inverse speed
\begin{equation}
  \eta\!\left(\vmin\right) \;=\;
  \int_{v \ge \vmin} d^{3}v\;\frac{f_{\chi}(\bm{v})}{v}\,.
  \label{eq:eta}
\end{equation}
Two well-known kinematic facts~\cite{essig2012,graham2012,catena2020} shape the signal. First, in this intrinsically inelastic process the entire DM kinetic energy is in principle available for deposition, so the ionization spectra terminate at the endpoint $\Ee^{\mathrm{max}}\simeq\tfrac{1}{2}\mchi v_{\mathrm{max}}^{2}-E_{B}$, with $v_{\mathrm{max}}=v_{E}+v_{\mathrm{esc}}\simeq788~\mathrm{km\,s^{-1}}$, even $\mchi\sim\mathcal{O}(10~\mathrm{MeV})$ particles can ionize the outer shells of carbon and fluorine ($E_{B}^{2p}\simeq12$--$20~\mathrm{eV}$, Table~\ref{tab:orbitals}). Second, depositing $\Delta E$ requires $q\gtrsim\Delta E/v_{\mathrm{max}}$, far more than a free electron at rest could absorb. The rate is therefore controlled by the high-momentum tail of the bound-state wavefunctions, encoded in a steeply falling ionization form factor $|\fion(k',q)|^{2}$, with $k'=\sqrt{2\me\Ee}$ the momentum of the outgoing electron.

Following the standard formalism~\cite{essig2012,essig2017,catena2020}, the differential ionization rate per unit target mass, summed over the occupied orbitals of the target, is
\begin{equation}
  \begin{split}
    \frac{dR}{d\ln \Ee}
    \;=\;{}& N_{T}\,\frac{\rhochi}{\mchi}\,
    \sum_{\mathrm{orb}}
    \frac{\sigmae}{8\,\mue^{2}} \\
    &\times \int dq\; q\,\bigl|\Fdm(q)\bigr|^{2}\,
    \bigl|\fion(k',q)\bigr|^{2}\,
    \eta\!\left(\vmin\right),
  \end{split}
  \label{eq:rate}
\end{equation}
with $dR/d\Ee=(1/\Ee)\,dR/d\ln\Ee$. Here $\mue$ is the DM--electron reduced mass, $N_{T}=N_{\mathrm{A}}/M_{\mathrm{mol}}\simeq5.90\times10^{24}~\mathrm{kg^{-1}}$ is the number of \CHF{} units per unit target mass ($M_{\mathrm{mol}}=102.03~\mathrm{g\,mol^{-1}}$), and $\sigmae$ is the reference DM--electron cross section, defined from the free scattering amplitude evaluated at the reference momentum transfer $q_{\mathrm{ref}}=\aem\me\simeq3.7~\mathrm{keV}$, the typical momentum of an outer-shell electron; see Refs.~\cite{essig2012,essig2016semic,catena2020} for the explicit construction. The residual $q$-dependence of the amplitude is carried by the DM form factor,
\begin{equation}
  \Fdm(q) \;=\;
  \begin{cases}
    1, & \text{heavy mediator,}\\[4pt]
    \left(\dfrac{\aem\me}{q}\right)^{2}, & \text{light mediator,}
  \end{cases}
  \label{eq:FDM}
\end{equation}
the two limiting cases of the dark-photon benchmark~\cite{Holdom:1985ag,essig2012}: a mediator heavy compared to the momentum transfer yields the contact interaction $\Fdm=1$, while an ultralight mediator yields the long-range interaction $\Fdm=(\aem\me/q)^{2}$. In the general non-relativistic effective theory of DM--electron interactions~\cite{catena2020}, this benchmark -- and any interaction dominated by the leading operator $\mathcal{O}_{1}=1_{\chi}1_{e}$ -- probes only the first atomic response function, $W_{1}=|\fion|^{2}$, i.e., the standard ionization form factor entering Eq.~\eqref{eq:rate}. The additional responses $W_{2,3,4}$, which arise for anapole-, magnetic-dipole-, and electric-dipole--type couplings, are not probed here, and we refer to Refs.~\cite{catena2020,catena2023beyond,liang2024} for the general treatment. We present all results for both mediator limits.

\subsection{Ionization form factor of the \CHF{} target}\label{sec:ff}
The target-specific input to Eq.~\eqref{eq:rate} is the ionization form factor of \CHF. We evaluate it in the \emph{isolated-atom superposition} approximation i.e., the target is treated as an incoherent superposition of its constituent atoms, and the ionization response is the stoichiometry-weighted sum of atomic ones,
\begin{equation}
  \bigl|\fion^{\CHF}(k',q)\bigr|^{2}
  =\!\!\sum_{A\,\in\,\{\mathrm{C,H,F}\}}\!\!\nu_{A}
  \sum_{(n\ell)\,\in\, A}\bigl|\fion^{n\ell,A}(k',q)\bigr|^{2},
  \label{eq:fion-mol}
\end{equation}
with multiplicities $\nu_{\mathrm{C}}=\nu_{\mathrm{H}}=2$ and $\nu_{\mathrm{F}}=4$, each orbital entering Eq.~\eqref{eq:vmin} with its own binding energy (Table~\ref{tab:orbitals}). This approximation is justified by the kinematics. The signal is dominated by momentum transfers $q\gtrsim\aem\me\simeq3.7~\mathrm{keV}$~\cite{liang2024}, i.e., by length scales well below the Bohr radius, where the electron wavefunctions are governed by the atomic potentials and are only weakly distorted by the $\mathcal{O}(\mathrm{eV})$ rearrangements associated with chemical bonding. An analogous isolated-atom treatment underlies the standard analyses of liquid noble-gas detectors~\cite{essig2012,essig2017,catena2020}. It is least accurate for the valence electrons participating in the C--H, C--C, and C--F bonds, whose binding energies and momentum distributions are modified at the $\mathcal{O}(10~\mathrm{eV})$ level. The dedicated calculations of the kind now available for aromatic organic targets~\cite{Blanco:2019lrf} are the natural refinement and are left to future work.

For the bound states of carbon and fluorine, we employ the Roothaan--Hartree--Fock ground-state wavefunctions of Ref.~\cite{Bunge:1993jsz}, with binding energies given by the corresponding eigenvalues, and the hydrogen $1s$ orbital is treated exactly. The outgoing electron is described by a positive-energy continuum solution of the Schr\"odinger equation in the hydrogenic potential of the residual ion, with effective charge $Z_{\mathrm{eff}}^{n\ell}=n\,\sqrt{E_{B}^{n\ell}/13.6~\mathrm{eV}}$ fixed by the binding energy of the ionized orbital and normalization as in Ref.~\cite{catena2020}. All bound--continuum overlap integrals are evaluated numerically with the publicly available \texttt{DarkART} code~\cite{DarkART}, which implements the formalism of Ref.~\cite{catena2020}. We validated our setup by reproducing the published argon and xenon ionization responses of Ref.~\cite{catena2020} before applying it to the \CHF{} constituents. The explicit radial forms and normalization conventions, together with the orbitals included in the analysis, their binding energies, occupancies, and effective charges, are collected in Appendix~\ref{app:estructure} (Table~\ref{tab:orbitals}).

\subsection{From event rate to projected limits}\label{sec:limits}
For a given DM mass and mediator type, the expected number of signal events is
\begin{equation}
  N_{\mathrm{exp}}(\mchi,\sigmae)
  \;=\;\varepsilon\,{\cal E}
  \int_{E_{\mathrm{th}}}^{\Ee^{\mathrm{max}}} d\Ee\;
  \frac{dR}{d\Ee}\,,
  \label{eq:Nexp}
\end{equation}
where ${\cal E}$ is the exposure (target mass $\times$ live time), $\varepsilon$ the detection efficiency (unity above threshold; Sec.~\ref{sec:detector}), and $E_{\mathrm{th}}$ the Seitz threshold at the corresponding operating temperature (Table~\ref{tab:exclusion}). We assume a benchmark exposure of ${\cal E}=10^{3}~\mathrm{kg\,days}$ throughout, of the order of the exposures envisaged for forthcoming \InDEx{} physics runs~\cite{Kumar:2025ofs,das2025index}. Since no background model is assumed (Sec.~\ref{sec:detector}) and the rate is linear in $\sigmae$, the $90\%$~C.L.\ upper limit follows from the standard zero-background, zero-observed-event Poisson condition $N_{\mathrm{exp}}\leq2.3$~\cite{seth2020,baxter2021},
\begin{equation}
  \sigmae^{90}(\mchi)
  \;=\;\frac{2.3}{\varepsilon\,{\cal E}}
  \left[\;\int_{E_{\mathrm{th}}}^{\Ee^{\mathrm{max}}} d\Ee\;
  \frac{1}{\sigmae}\frac{dR}{d\Ee}\right]^{-1},
  \label{eq:limit}
\end{equation}
where the bracket is $\sigmae$-independent.

\begin{figure*}[!t]
    \centering
    \includegraphics[width=0.48\textwidth]{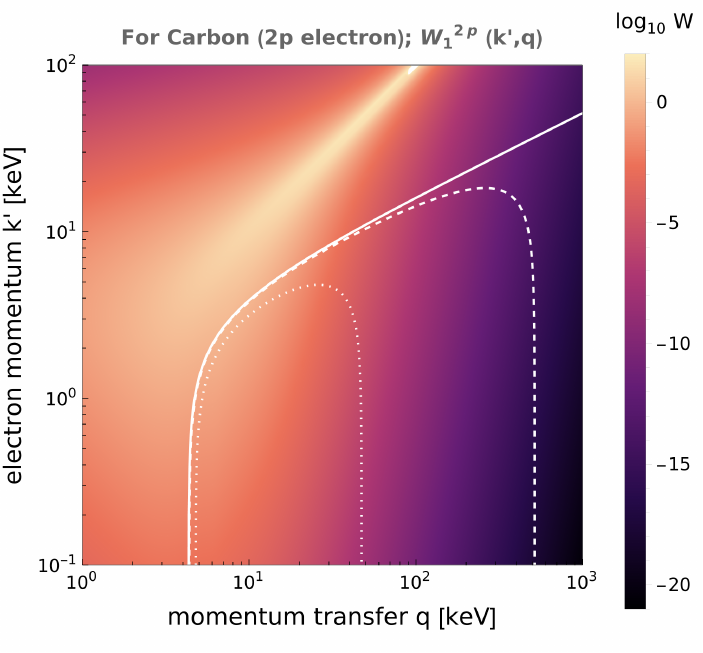}
    \hfill
    \includegraphics[width=0.48\textwidth]{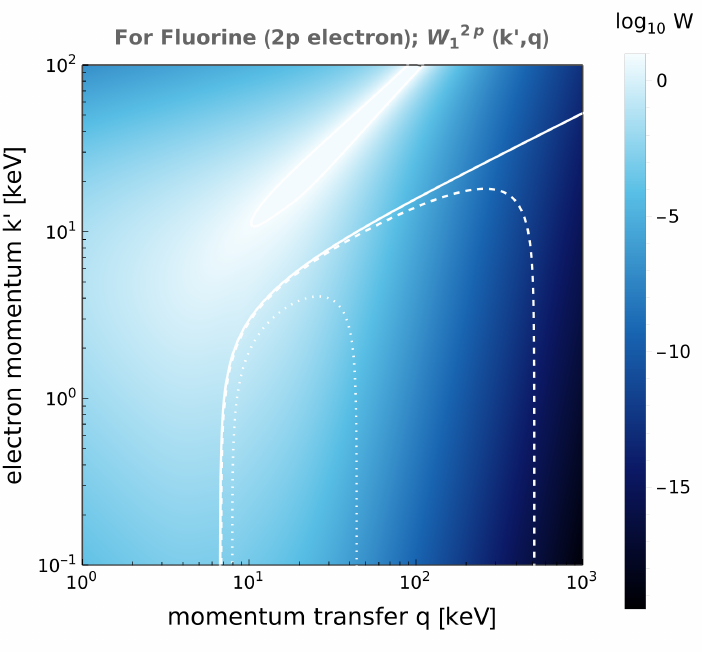}
    \caption{Ionization response $W_1^{2p}(k',q)=|\fion^{2p}(k',q)|^{2}$ for the
    $2p$ electrons of carbon (left panel) and fluorine (right panel), the
    dominant light constituents of the C$_2$H$_2$F$_4$ target, as calculated
    using the \texttt{DarkART} package~\cite{DarkART}. The response,
    shown as $\log_{10} W_1$, is displayed in the plane of the momentum
    transfer $q$ and the asymptotic momentum $k'$ of the outgoing electron,
    following the formalism of Ref.~\cite{catena2020}. The white curves mark
    the kinematic boundary imposed by the requirement
    $v_\mathrm{min}(k',q) < v_\mathrm{max} = v_\oplus + v_\mathrm{esc}$:
    above each curve, the minimum dark matter speed needed to ionize the
    orbital exceeds the maximum speed available in the galactic halo, so the
    corresponding region of $(q,k')$ space cannot be probed. The dotted,
    dashed, and solid curves correspond to dark matter masses
    $m_\chi = 10~\mathrm{MeV}$, $100~\mathrm{MeV}$, and $m_\chi \to \infty$,
    respectively.}
    \label{fig:W1_response}
\end{figure*}

\section{Results and discussion}\label{sec:results}

\subsection{Ionization response of the \CHF{} target}\label{sec:res-response}
Figure~\ref{fig:W1_response} shows the ionization response $W_{1}^{2p}=|\fion^{2p}|^{2}$ of the carbon (left) and fluorine (right) $2p$ orbitals, which dominate the low-threshold ionization signal of the \CHF{} target. Both responses are largest along a ridge at momentum transfers of a few keV, of the order of the characteristic bound-state momentum $Z_{\mathrm{eff}}\,\aem\me$, and fall steeply towards large $q$ and large $k'$, reflecting the decreasing support of the bound-state momentum distribution at high momenta. At fixed $k'$, the fluorine response extends to visibly larger momentum transfers than that of carbon, a consequence of its larger effective nuclear charge and more compact orbital. This harder tail, together with the four-fold multiplicity of fluorine, anticipates the dominance of fluorine in the target spectra discussed below. The superimposed white curves delimit the kinematically accessible region for a given $\mchi$ [Eq.~\eqref{eq:vmin}]: for $\mchi=10~\mathrm{MeV}$ only a small low-$q$, low-$k'$ corner of the response is accessible, while for heavier DM essentially the full ridge opens up -- the kinematic origin of the rapid growth of the total rate with $\mchi$ near the low-mass edge seen in Figs.~\ref{fig:rates} and \ref{fig:rates_LM}.

\begin{figure*}[!t]
  \centering
  \includegraphics[width=0.98\textwidth]{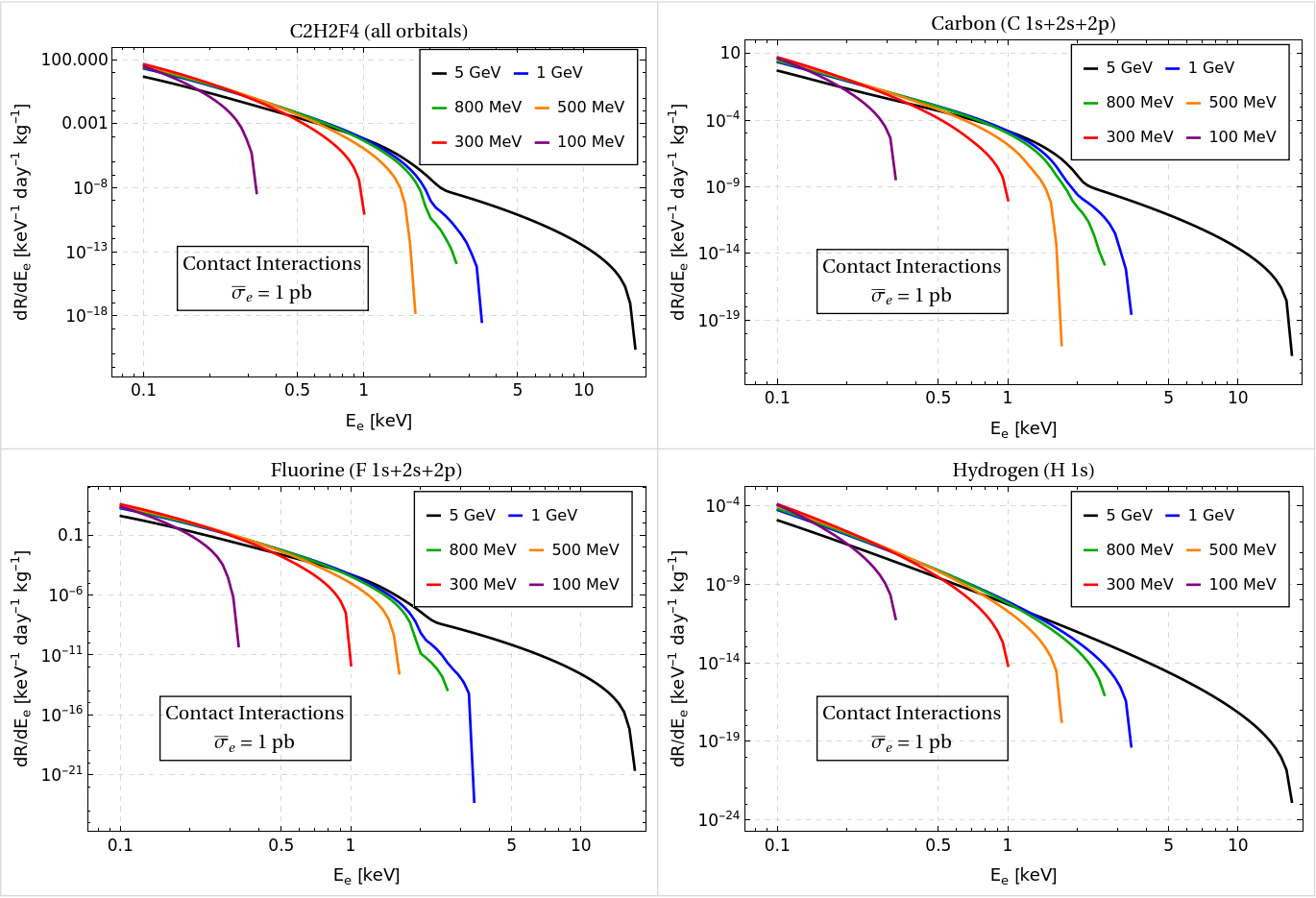}
  \caption{%
    Differential electron-recoil spectra $dR/dE_e$ for the C$_2$H$_2$F$_4$ (R-134a) target, assuming a contact interaction ($F_{\rm DM}(q)=1$, i.e.\ a heavy mediator) and a reference DM--electron cross section $\bar\sigma_e=1~\mathrm{pb}$. Panels show the spectrum summed over the full target (top left) and its decomposition into carbon $1s{+}2s{+}2p$ (top right), fluorine $1s{+}2s{+}2p$ (bottom left), and hydrogen $1s$ (bottom right). Within each panel, the six curves correspond to DM masses $m_\chi = 5,\,1,\,0.8,\,0.5,\,0.3,\,0.1~\mathrm{GeV}$.}
  \label{fig:recoil}
\end{figure*}

\subsection{Ionization spectra and event rates}\label{sec:res-rates}
Figure~\ref{fig:recoil} shows the differential ionization spectra $dR/dE_e$ for a contact interaction ($F_{\rm DM}=1$) at $\bar\sigma_e=1~\mathrm{pb}$. The spectra fall steeply with recoil energy and terminate at the kinematic endpoint, $E_e^{\rm max}\simeq \tfrac{1}{2}m_\chi v_{\rm max}^2-E_B$, set by the maximum DM speed in the halo, such that the accessible recoil-energy range increases with $m_\chi$. For the benchmark masses considered here, the endpoint extends from $\lesssim0.3~\mathrm{keV}$ at $m_\chi=100~\mathrm{MeV}$ to $\simeq17~\mathrm{keV}$ at $m_\chi=5~\mathrm{GeV}$. At higher recoil energies, the spectra develop a distinct shoulder, reflecting the transition from the steeply falling contribution of the loosely bound valence electrons to that of the more deeply bound inner shells, particularly the $1s$ states of carbon and fluorine. The compact inner-shell wavefunctions contain larger high-momentum components and consequently give rise to a harder recoil tail. Hydrogen, with only a single $1s$ state, does not exhibit a comparable feature. The species decomposition further shows that fluorine provides the dominant contribution to the signal, with carbon subdominant and hydrogen negligible, consistent with the four fluorine atoms per \CHF{} molecule and their larger effective electron content. Throughout, the standard halo model is assumed, with $\rho_\chi=0.4~\mathrm{GeV\,cm^{-3}}$, $v_0=220~\mathrm{km\,s^{-1}}$, $v_{\rm esc}=544~\mathrm{km\,s^{-1}}$, and $v_E=244~\mathrm{km\,s^{-1}}$.

\begin{figure*}[!t]
  \centering
  \includegraphics[width=0.98\textwidth]{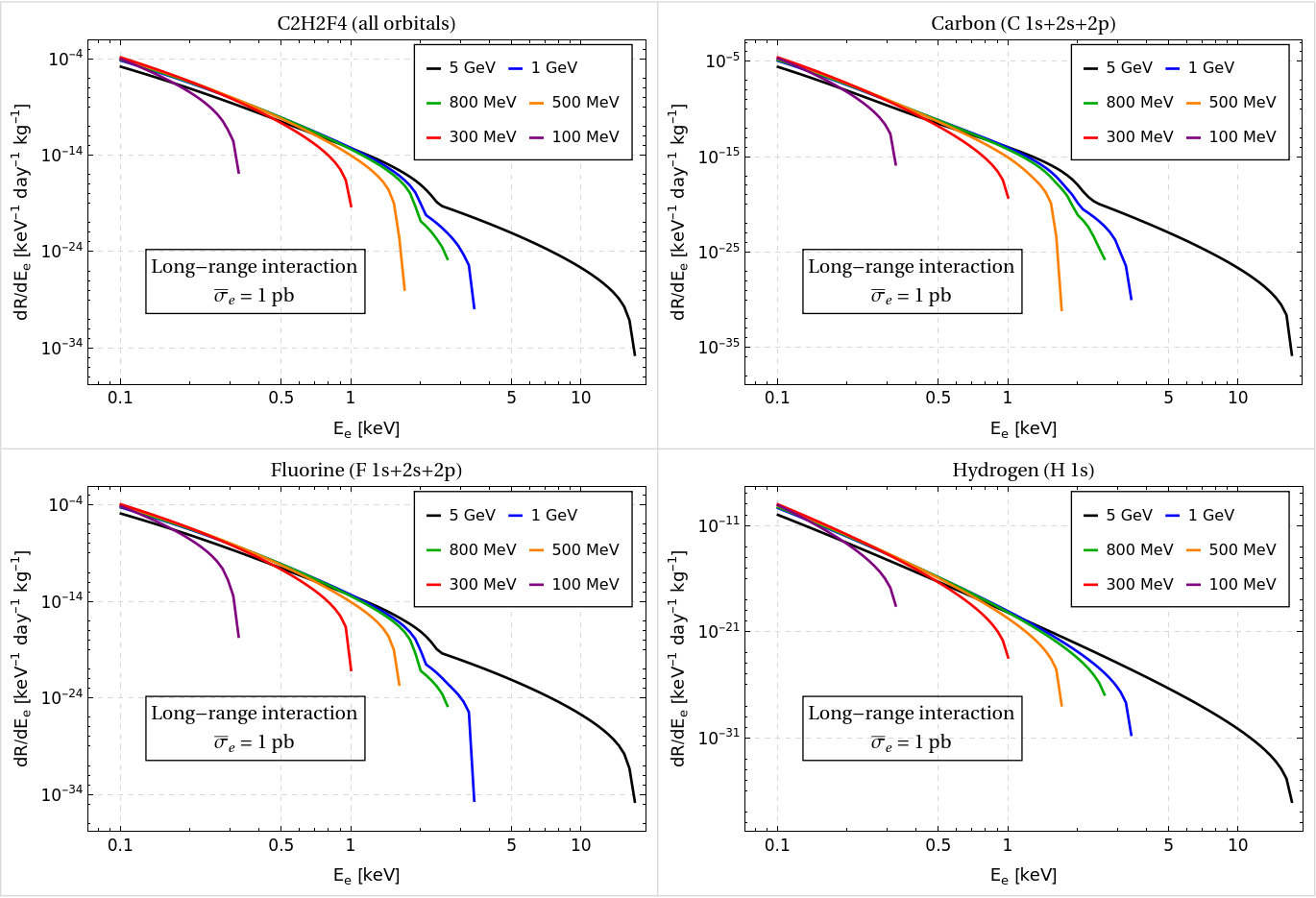}
  \caption{%
    Differential electron-recoil spectra $dR/dE_e$ for the C$_2$H$_2$F$_4$ (R-134a) target, assuming a \emph{long-range} interaction---a light/ultralight mediator with
    $F_{\rm DM}(q)=(\alpha m_e/q)^2$---and a reference DM--electron cross section $\bar\sigma_e=1~\mathrm{pb}$. Panels show the spectrum summed over the full target (top left) and its decomposition into carbon $1s{+}2s{+}2p$ (top right), fluorine $1s{+}2s{+}2p$ (bottom left), and hydrogen $1s$ (bottom right); within each panel the six curves correspond to $m_\chi = 5,\,1,\,0.8,\,0.5,\,0.3,\,0.1~\mathrm{GeV}$.}
  \label{fig:recoil_LM}
\end{figure*}

Figure~\ref{fig:recoil_LM} shows the corresponding differential ionization spectra for the long-range interaction, $F_{\rm DM}(q)=(\alpha m_e/q)^2$. The kinematic endpoints are independent of the mediator and coincide with those of the contact interaction, while the normalization and spectral shape differ substantially. For a deposited energy $\Delta E$, the momentum transfer is bounded from below by $q\gtrsim\Delta E/v_{\rm max}$, which lies well above the reference momentum $q_{\rm ref}=\alpha m_e$ used to define $\bar\sigma_e$. Consequently, the $|F_{\rm DM}(q)|^2=(\alpha m_e/q)^4$ factor strongly suppresses the long-range spectra at fixed $\bar\sigma_e$, with the suppression becoming increasingly pronounced at higher recoil energies and leading to a steeper spectral fall-off. The same $q^{-4}$ weighting also concentrates the momentum-transfer integral in Eq.~\eqref{eq:rate} towards its lower kinematic limit, which is resolved using a logarithmic $q$ grid. The species hierarchy remains unchanged, with fluorine providing the dominant contribution, carbon subdominant, and hydrogen negligible.

\begin{figure*}[!t]
  \centering
  \begin{minipage}{0.48\textwidth}
    \centering
    \includegraphics[width=\linewidth]{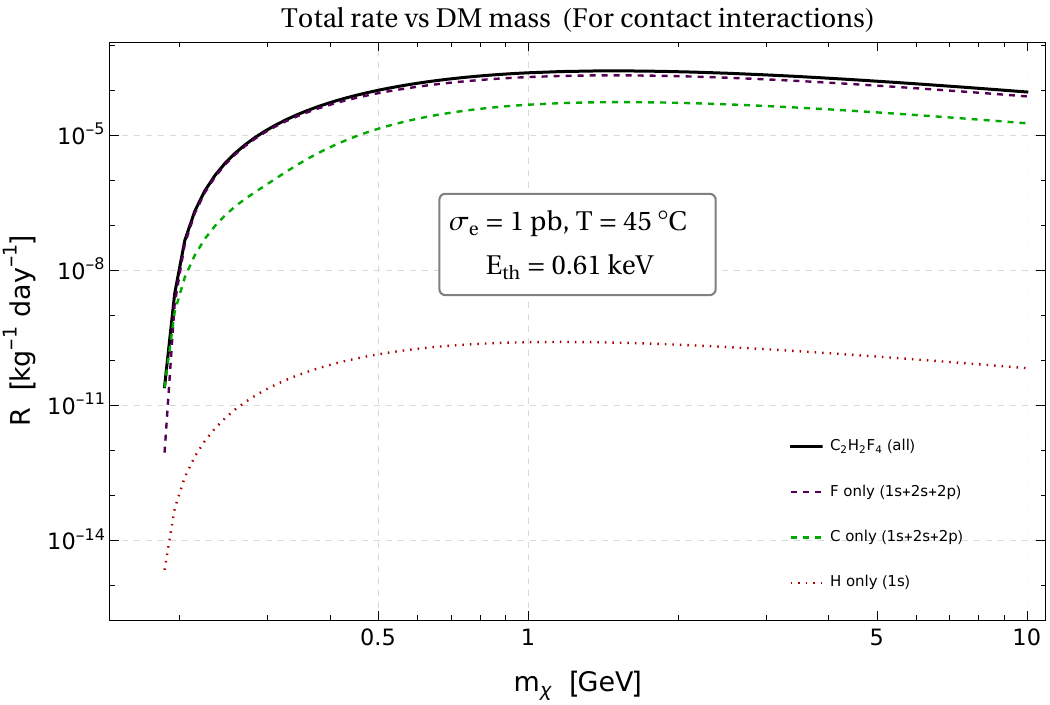}\\[2pt]
    {\small (a) 
    }
  \end{minipage}\hfill
  \begin{minipage}{0.48\textwidth}
    \centering
    \includegraphics[width=\linewidth]{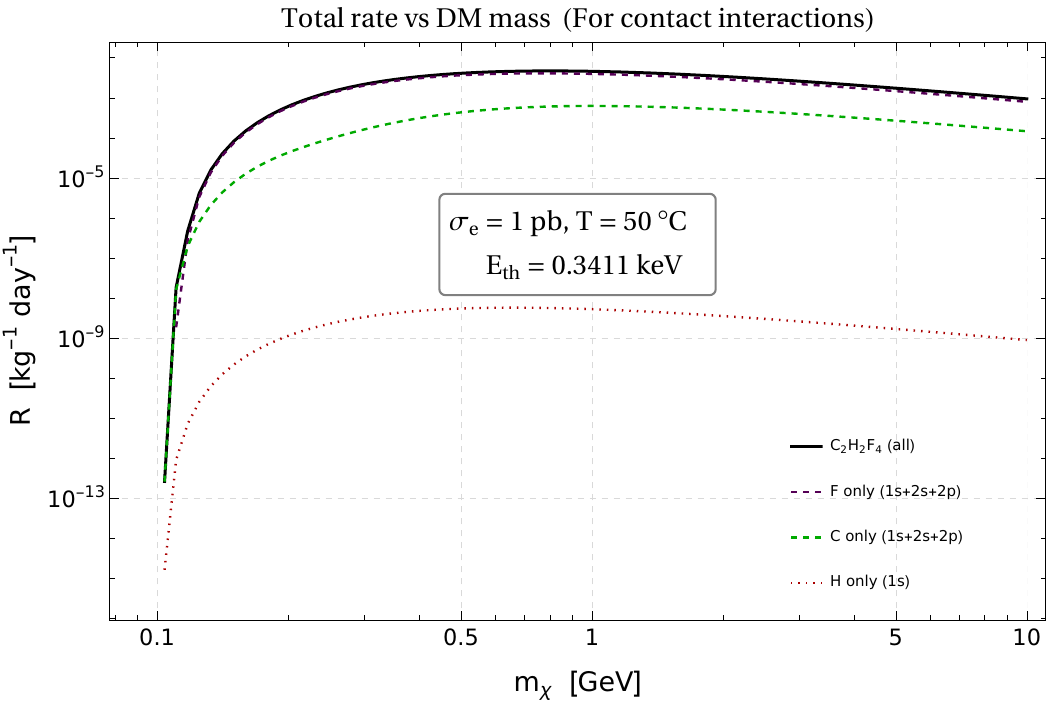}\\[2pt]
    {\small (b) 
    }
  \end{minipage}

  \vspace{0.8em}
  \begin{minipage}{0.48\textwidth}
    \centering
    \includegraphics[width=\linewidth]{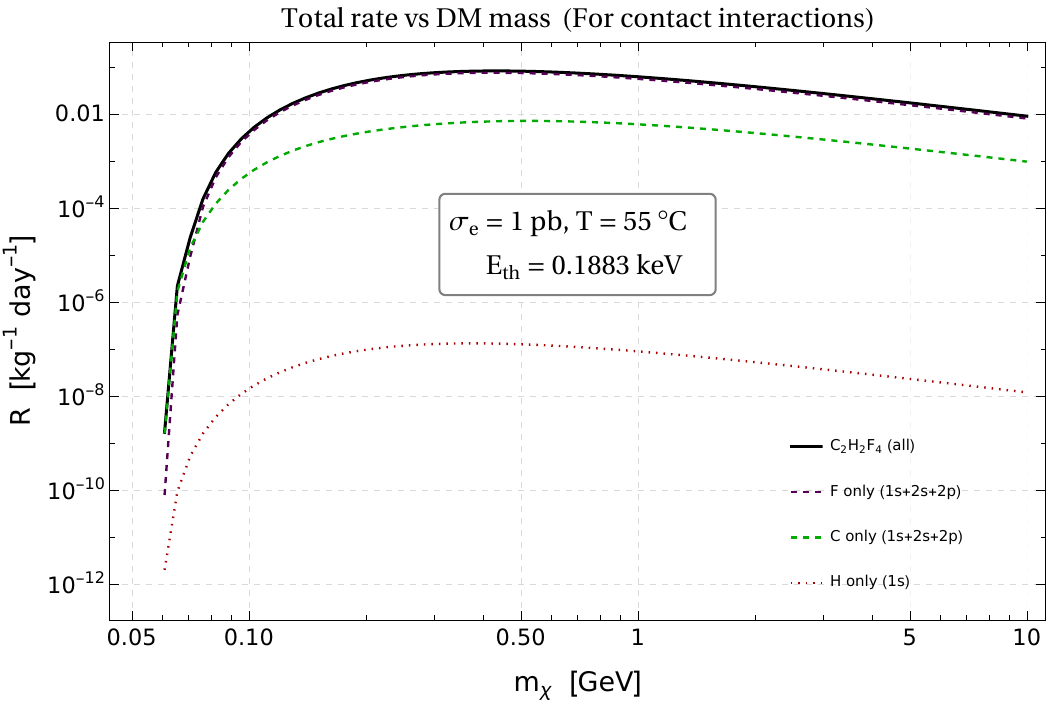}\\[2pt]
    {\small (c) 
    }
  \end{minipage}\hfill
  \begin{minipage}{0.48\textwidth}
    \centering
    \includegraphics[width=\linewidth]{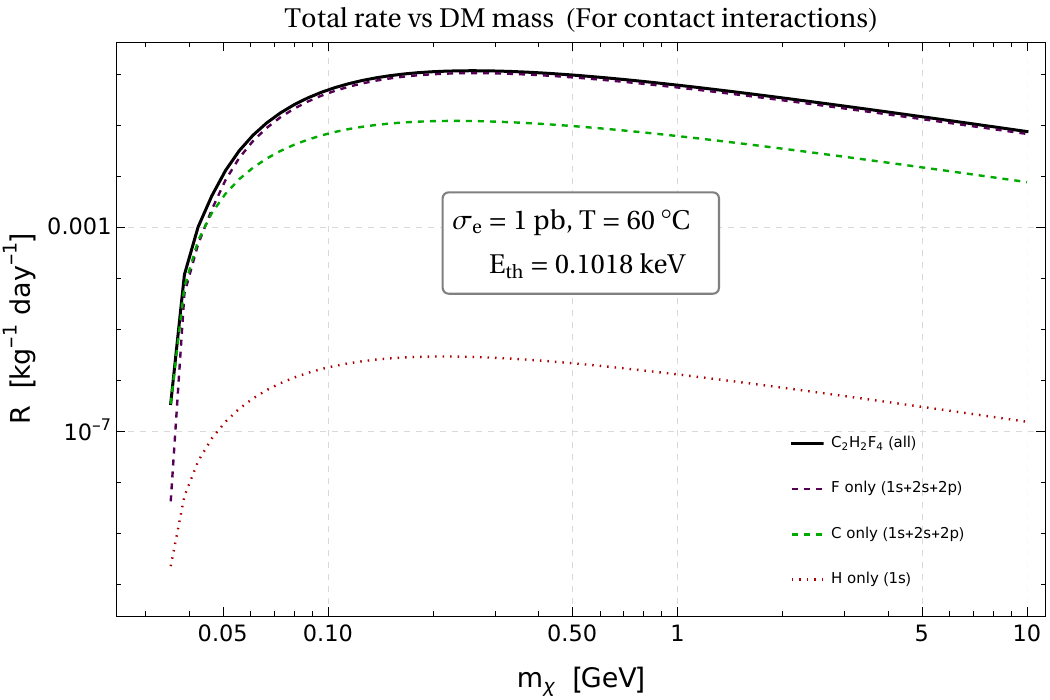}\\[2pt]
    {\small (d) 
    }
  \end{minipage}

  \vspace{0.8em}
  \begin{minipage}{0.48\textwidth}
    \centering
    \includegraphics[width=\linewidth]{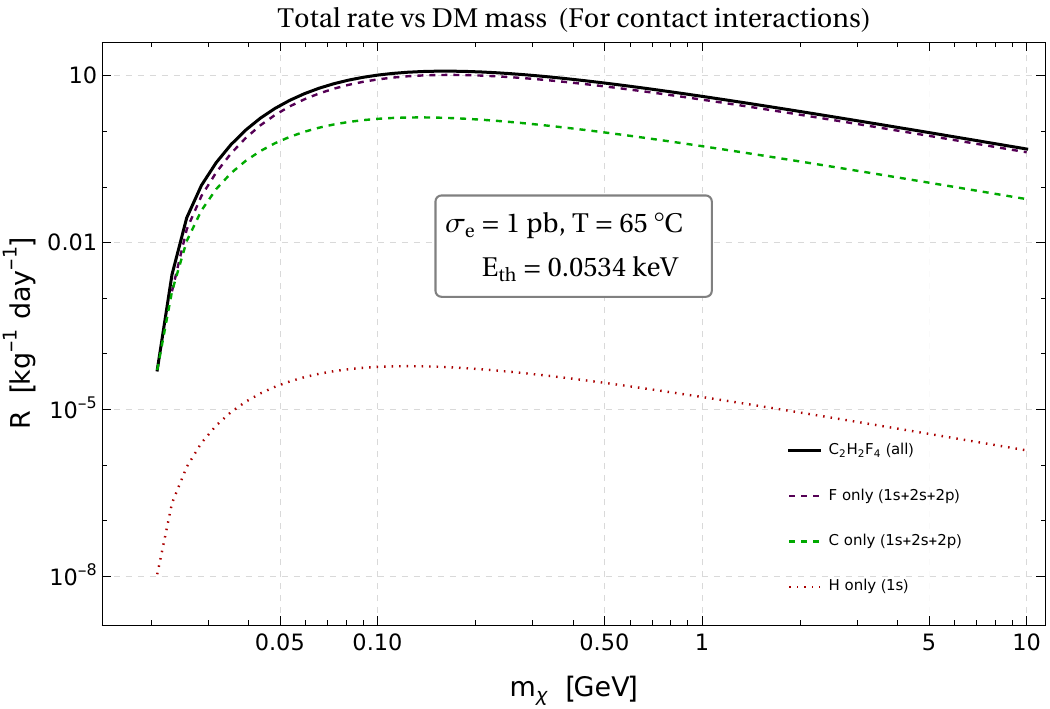}\\[2pt]
    {\small (e) 
    }
  \end{minipage}\hfill\begin{minipage}{0.48\textwidth}
    \centering
    \includegraphics[width=\linewidth]{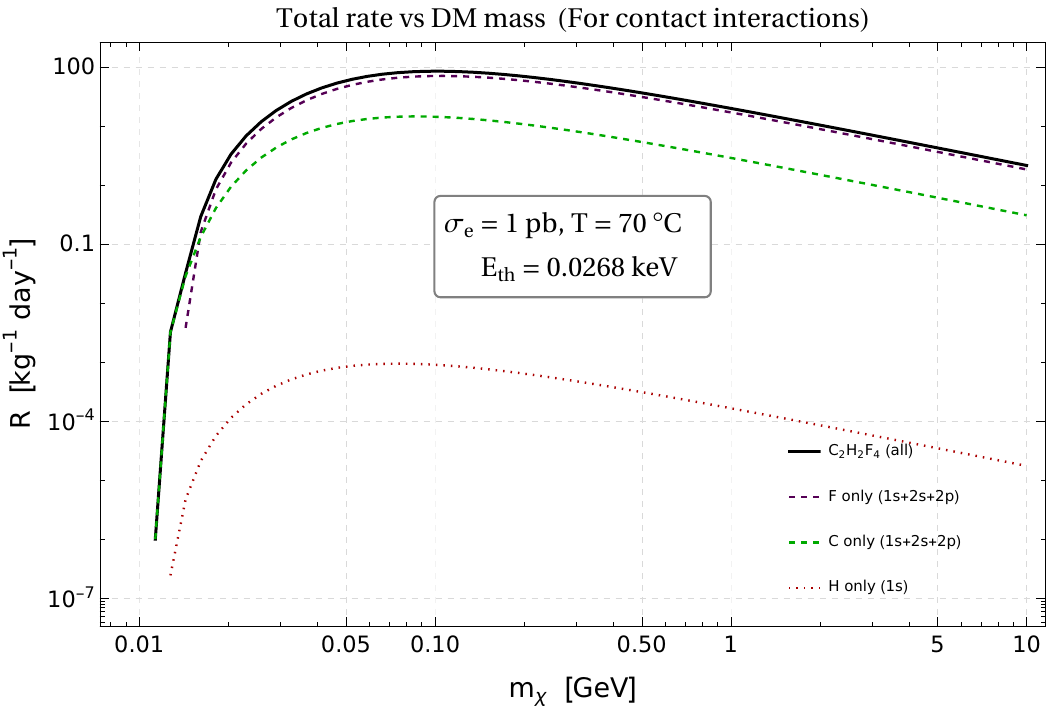}\\[2pt]
    {\small (f) 
    }
  \end{minipage}

  \caption{%
    Total DM--electron scattering rate $R$ per unit target mass versus DM mass $m_\chi$ for the C$_2$H$_2$F$_4$ target, for a contact interaction ($F_{\rm DM}=1$) at $\bar\sigma_e=1~\mathrm{pb}$. Panels (a)--(f) correspond to the six operating temperatures of the electron-channel window (Sec.~\ref{sec:detector}), with Seitz thresholds decreasing from $E_{\rm th}=0.61~\mathrm{keV}$ at $45~^\circ$C to $0.0268~\mathrm{keV}$ at $70~^\circ$C (cf.\ Table~\ref{tab:exclusion}). In each panel, the solid black curve represents the total rate of the target; the colored curves show the contributions from fluorine, carbon, and hydrogen.}
  \label{fig:rates}
\end{figure*}

\begin{figure*}[!t]
  \centering
  \begin{minipage}{0.48\textwidth}
    \centering
    \includegraphics[width=\linewidth]{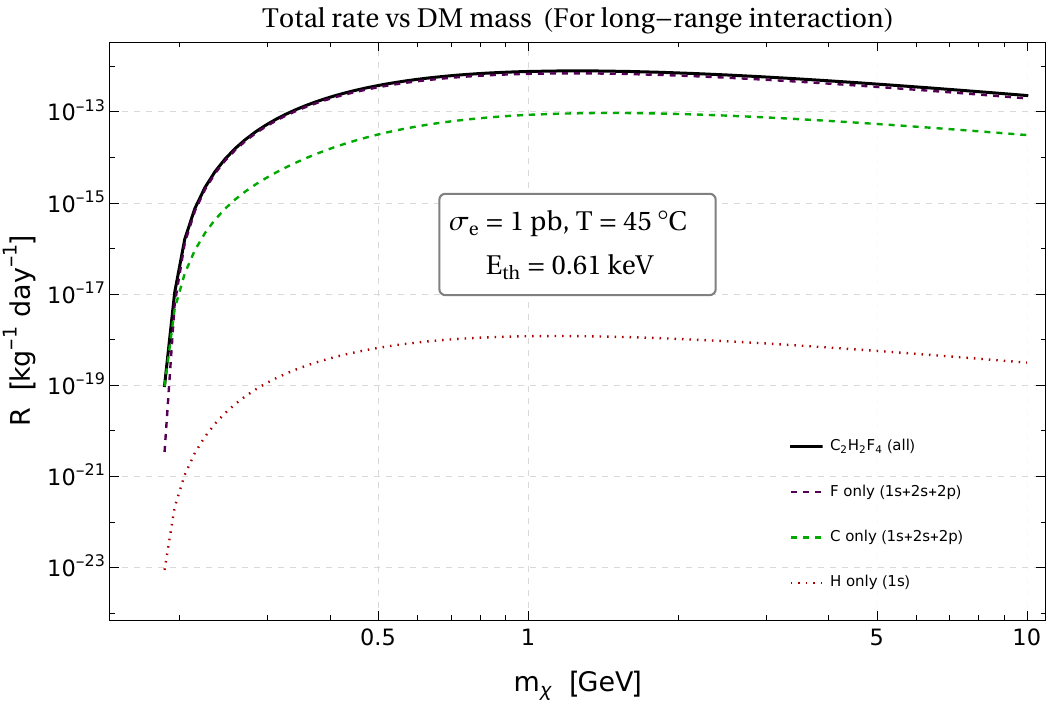}\\[2pt]
    {\small (a) 
    }
  \end{minipage}\hfill
  \begin{minipage}{0.48\textwidth}
    \centering
    \includegraphics[width=\linewidth]{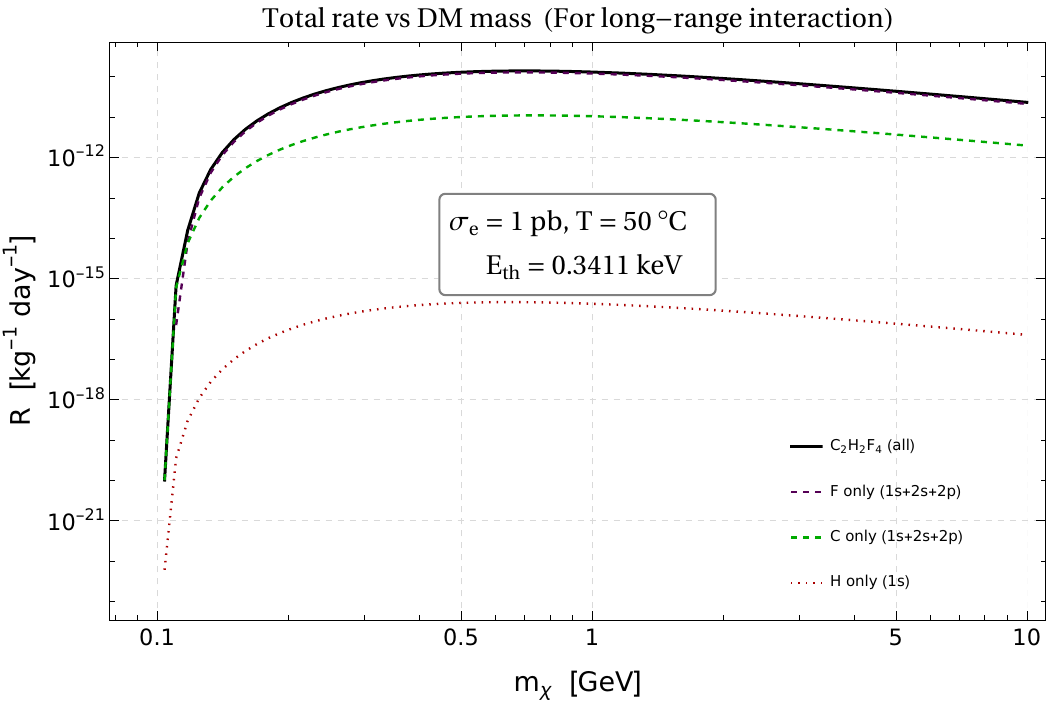}\\[2pt]
    {\small (b) 
    }
  \end{minipage}

  \vspace{0.8em}
  \begin{minipage}{0.48\textwidth}
    \centering
    \includegraphics[width=\linewidth]{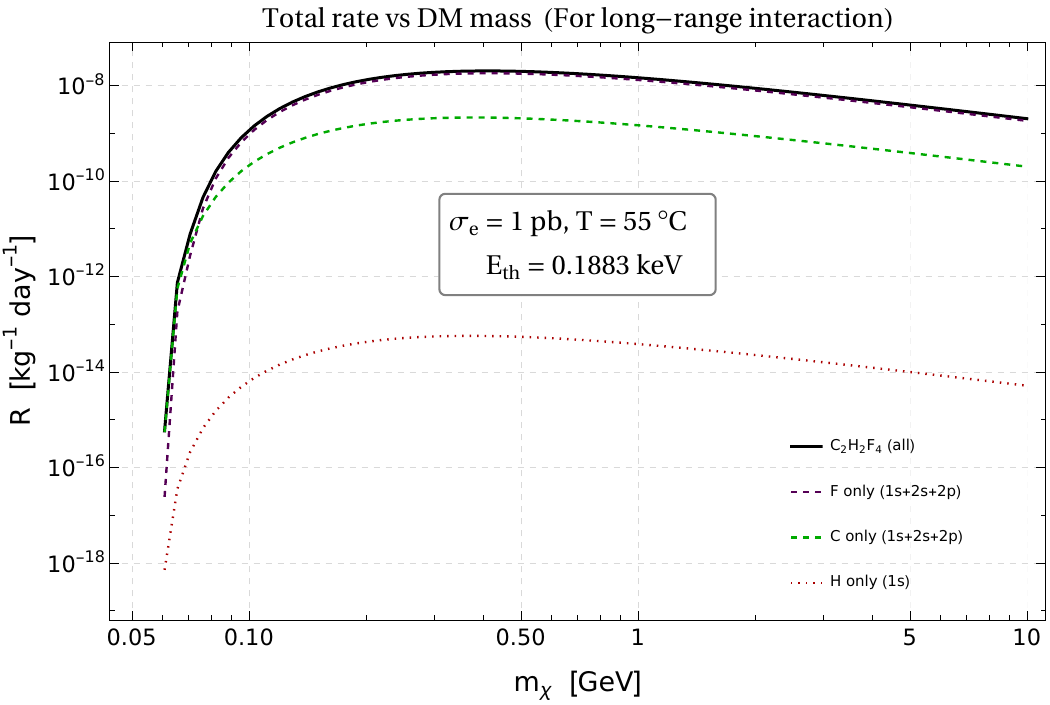}\\[2pt]
    {\small (c) 
    }
  \end{minipage}\hfill
  \begin{minipage}{0.48\textwidth}
    \centering
    \includegraphics[width=\linewidth]{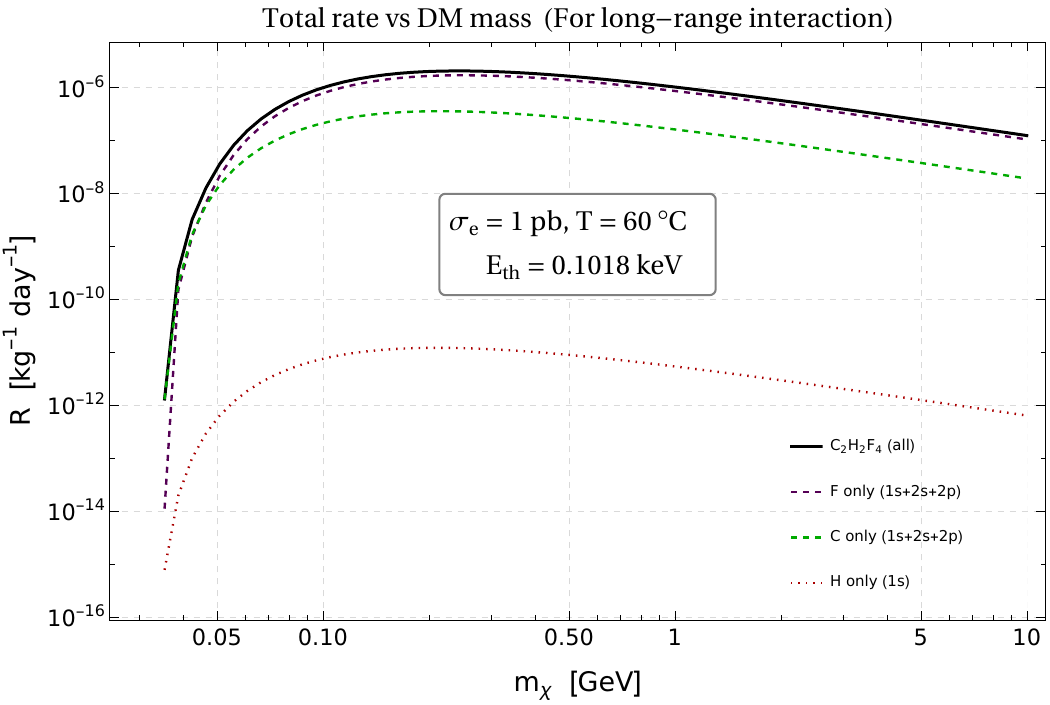}\\[2pt]
    {\small (d) 
    }
  \end{minipage}

  \vspace{0.8em}
  \begin{minipage}{0.48\textwidth}
    \centering
    \includegraphics[width=\linewidth]{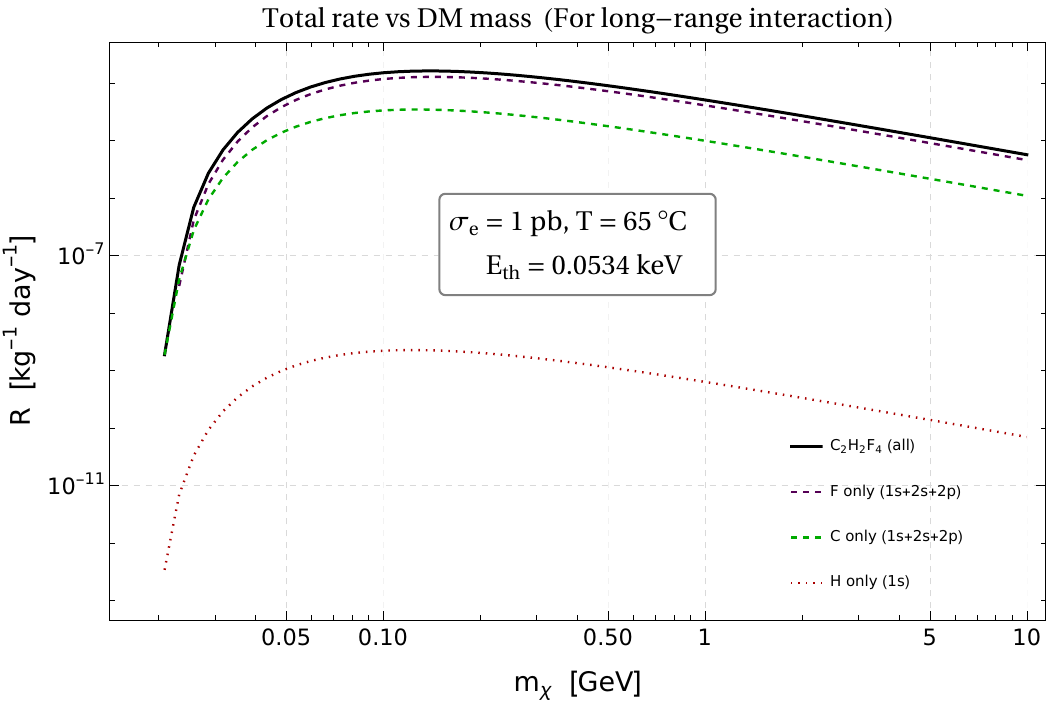}\\[2pt]
    {\small (e) 
    }
  \end{minipage}\hfill
 \begin{minipage}{0.48\textwidth}
    \centering
    \includegraphics[width=\linewidth]{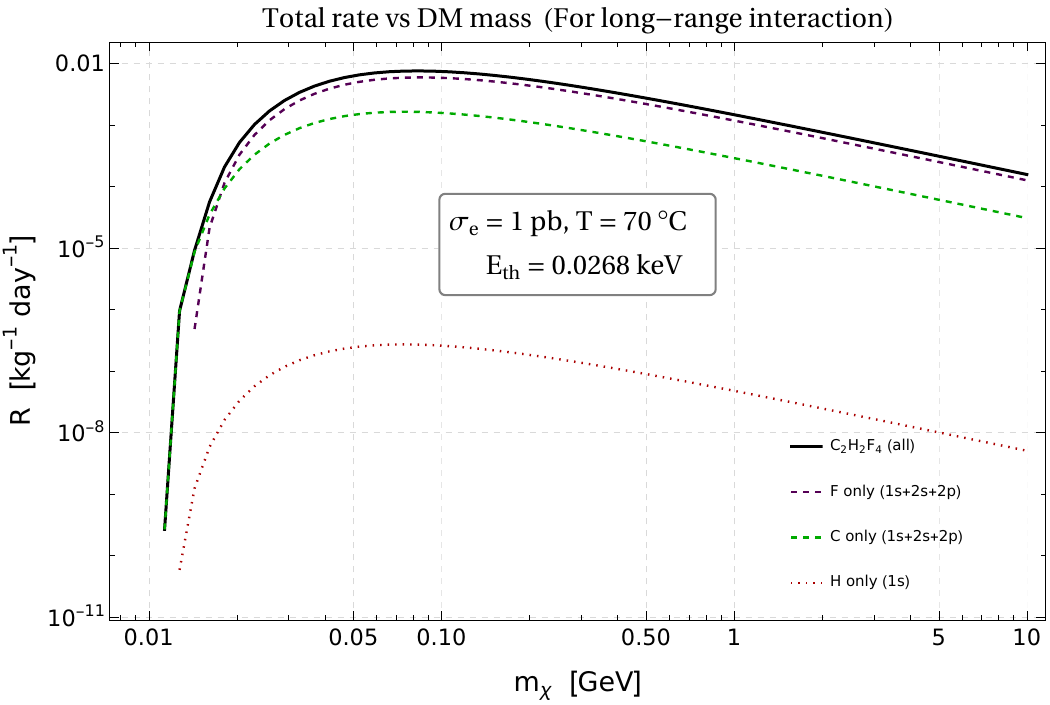}\\[2pt]
    {\small (f) 
    }
  \end{minipage}

  \caption{%
    Total DM--electron scattering rate $R$ per unit target mass versus DM mass $m_\chi$ for the C$_2$H$_2$F$_4$ target, for a long-range interaction ($F_{\rm DM}(q)=(\alpha m_e/q)^2$) at $\bar\sigma_e=1~\mathrm{pb}$. Panels (a)--(f) correspond to the six operating temperatures of the electron-channel window (Sec.~\ref{sec:detector}), with Seitz thresholds decreasing from $E_{\rm th}=0.61~\mathrm{keV}$ at $45~^\circ$C to $0.0268~\mathrm{keV}$ at $70~^\circ$C (cf.\ Table~\ref{tab:exclusion}). In each panel, the solid black curve is the total rate of the target; the colored curves give the fluorine, carbon, and hydrogen contributions.}
  \label{fig:rates_LM}
\end{figure*}

Integrating the differential ionization spectra above the six detector thresholds yields the total event rates shown in Fig.~\ref{fig:rates} for the contact interaction and Fig.~\ref{fig:rates_LM} for the long-range interaction as functions of $m_\chi$. Because a recoil must deposit at least $E_{\rm th}$ to nucleate a bubble, each curve exhibits a sharp low-mass cutoff determined by the condition $\tfrac12 m_\chi v_{\rm max}^2 \gtrsim E_{\rm th}+E_B$. Lowering the threshold from $0.61~\mathrm{keV}$ at $45~^\circ$C to $26.8~\mathrm{eV}$ at $70~^\circ$C relaxes this kinematic requirement, shifting the low-mass reach from $m_\chi\simeq0.18~\mathrm{GeV}$ to $m_\chi\simeq0.012~\mathrm{GeV}$ and increasing the event rate at fixed mass by several orders of magnitude. The operating temperature, therefore, provides the primary handle on the experiment's low-mass sensitivity.

The long-range interaction exhibits the same low-mass cutoff, since the kinematic condition is independent of the mediator. Its overall normalization, however, is substantially reduced relative to the contact case. Because $\bar\sigma_e$ is defined at the reference momentum $q_{\rm ref}=\alpha m_e$, whereas the ionization signal is dominated by momentum transfers $q\gg\alpha m_e$, the form-factor weighting $|F_{\rm DM}(q)|^2=(\alpha m_e/q)^4$ suppresses the event rates at fixed $\bar\sigma_e$. This suppression also shifts the maxima of the rate curves to slightly lower DM masses compared with the contact interaction. In both scenarios, fluorine dominates the signal, carbon provides a subdominant contribution, and hydrogen remains negligible. The halo parameters are as defined in Sec.~\ref{sec:res-rates}. These rates scale linearly with $\bar\sigma_e$ and form the basis of the exclusion limits presented below.

\subsection{Projected exclusion limits}\label{sec:res-limits}
\begin{figure*}[!t]
  \centering
  \begin{minipage}{0.81\textwidth}
    \centering
    \includegraphics[width=\linewidth]{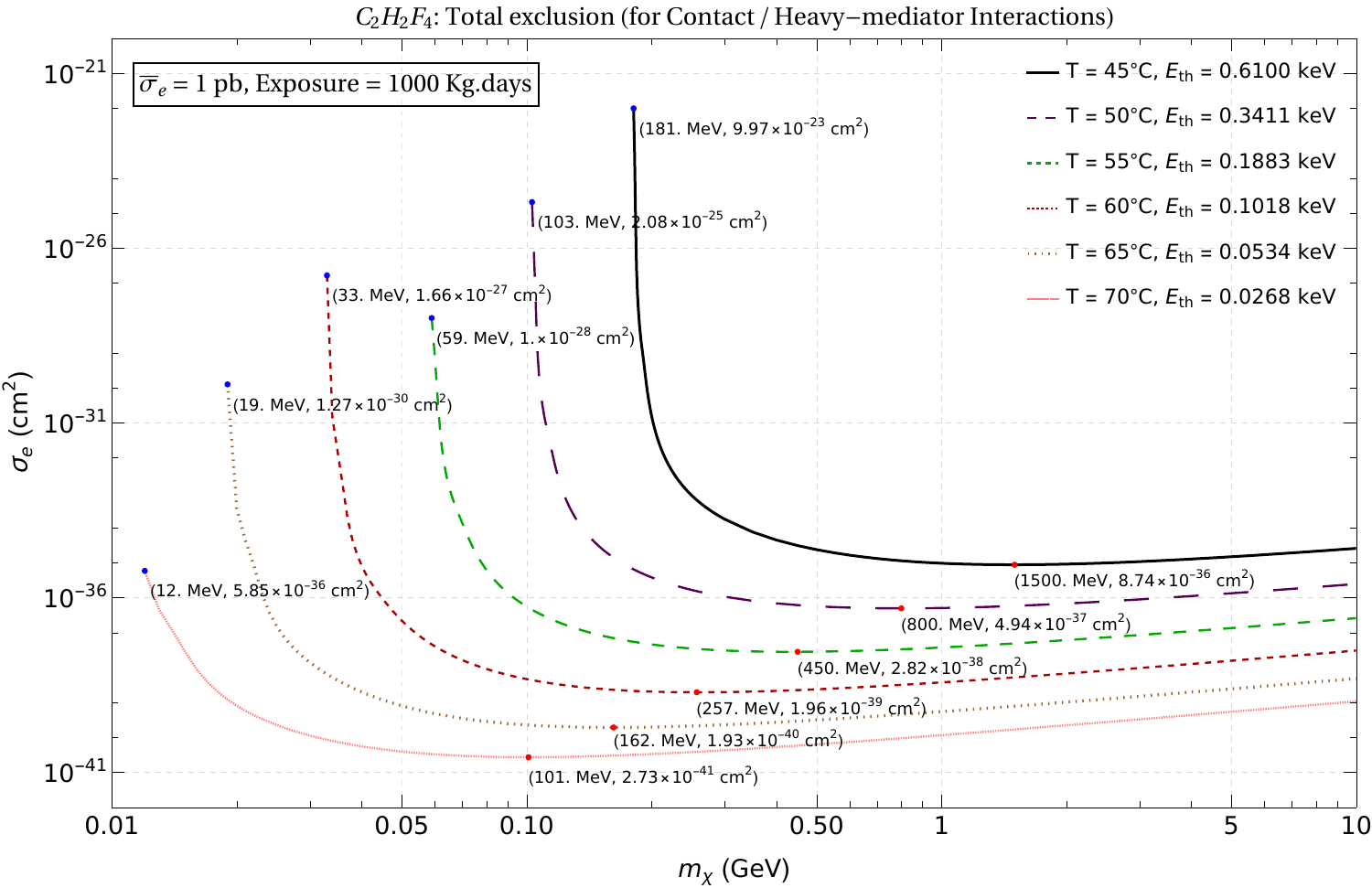}\\[2pt]
    {\small (a) }
  \end{minipage}  
  \begin{minipage}{0.81\textwidth}
    \centering
    \includegraphics[width=\linewidth]{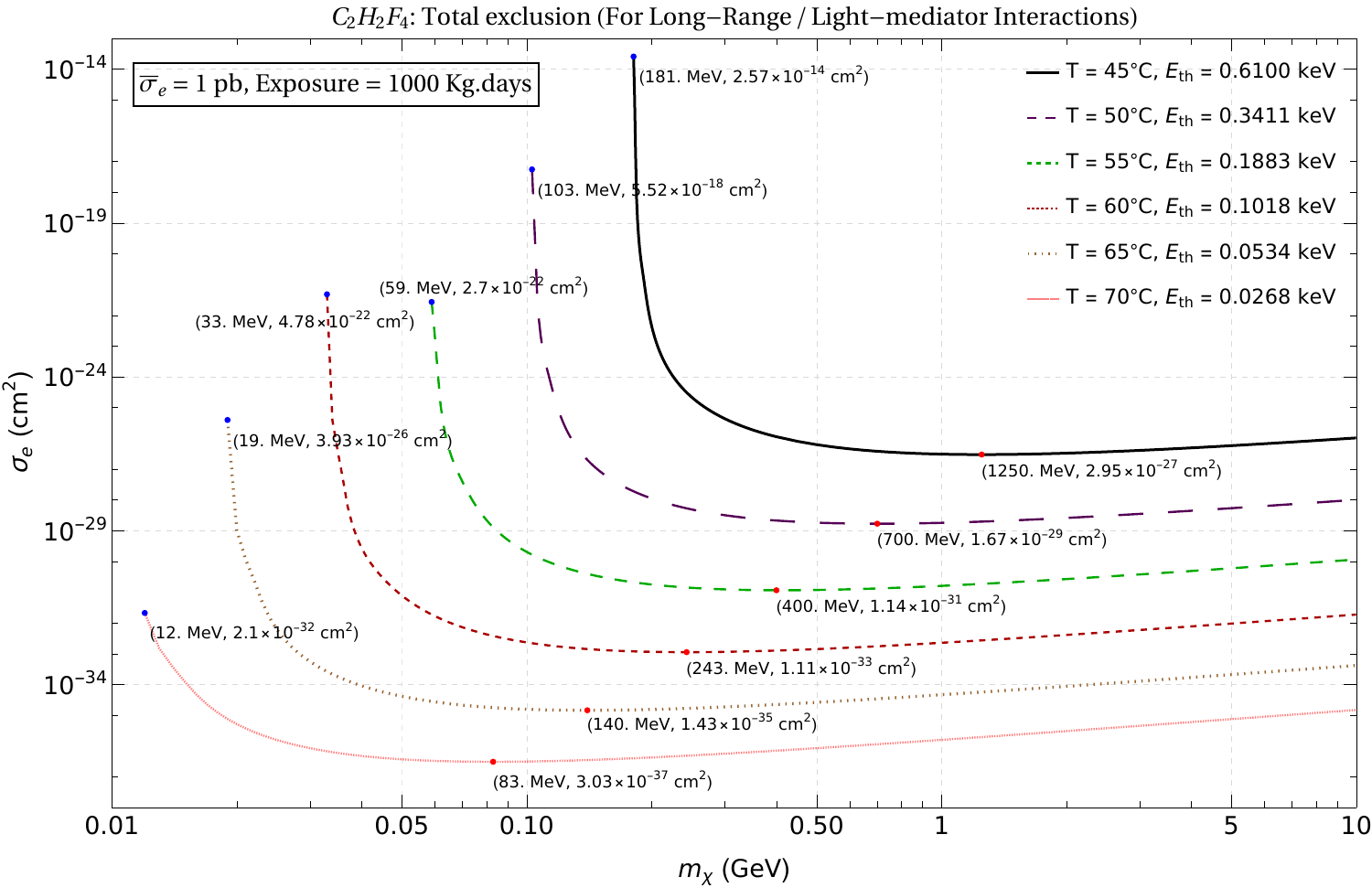}\\[2pt]
    {\small (b) }
  \end{minipage}
 \caption{Projected $90\%$~C.L.\ exclusion limits on the dark matter--electron scattering cross section $\sigmae$ as a function of the dark matter mass $\mchi$ for the \CHF{} target at different operating temperatures. The top panel~(a) shows the contact (heavy-mediator) case, $\Fdm(q)=1$, and the bottom panel~(b) shows the long-range (light-mediator) case, $\Fdm(q)=(\aem\me/q)^{2}$. The region above each curve is excluded. The limits assume a total exposure of $10^{3}~\mathrm{kg\, day}$ under the zero-background assumption of Sec.~\ref{sec:limits}. For each curve, the blue point marks the lowest dark matter mass that can be probed and the red point the most stringent cross-section limit; the annotated values are collected in Table~\ref{tab:exclusion}.}
  \label{fig:exclusion}
\end{figure*}

Figure~\ref{fig:exclusion} shows the projected $90\%$~C.L.\ exclusion limits on $\sigmae$ as a function of $\mchi$, obtained from Eq.~\eqref{eq:limit} at the six operating temperatures, for the contact ($\Fdm=1$, top panel) and long-range [$\Fdm=(\aem\me/q)^{2}$, bottom panel] interactions; they constitute the main result of this work. The corresponding low-mass edges and best sensitivities are collected in Table~\ref{tab:exclusion}. The limit curves share a characteristic shape. Towards low masses each curve rises steeply into a wall at the mass for which the maximal deposit $\tfrac12\mchi v_{\mathrm{max}}^{2}$ drops below $E_{\mathrm{th}}+E_{B}$, lowering the threshold from $0.61~\mathrm{keV}$ ($45~^{\circ}$C) to $26.8~\mathrm{eV}$ ($70~^{\circ}$C) moves this wall from $\mchi\simeq181~\mathrm{MeV}$ down to $\mchi\simeq12~\mathrm{MeV}$ and improves the limit at fixed mass by several orders of magnitude, confirming the operating temperature as the key experimental handle for this study. Towards high masses, the sensitivity degrades approximately linearly in $\mchi$, following the $\rhochi/\mchi$ decrease of the DM number density at fixed energy density. For the lowest threshold ($70~^{\circ}$C, $E_{\mathrm{th}}=26.8~\mathrm{eV}$), the contact-interaction limit reaches $\sigmae^{90}\simeq2.73\times10^{-41}~\mathrm{cm^{2}}$ near $\mchi\simeq101~\mathrm{MeV}$, while the long-range limit reaches $\sigmae^{90}\simeq3.03\times10^{-37}~\mathrm{cm^{2}}$ near $\mchi\simeq83~\mathrm{MeV}$; the difference in absolute scale reflects the normalization of $\sigmae$ at $q_{\mathrm{ref}}=\aem\me$. The lowest accessible mass is set by the kinematic wall and is therefore common to both mediator cases, while the $(\aem\me/q)^{4}$ weighting of the long-range rate, which favors the small momentum transfers characteristic of light DM, leaves the long-range curves comparatively stronger and flatter at the low-mass end.

\begin{table}[t]
\caption{Projected $90\%$~C.L.\ sensitivity of the \CHF{} DM--electron search ($10^{3}~\mathrm{kg\, day}$ exposure) at the 6 operating temperatures for both contact and long-range interactions. For every case, we list the kinematic low-mass edge $(\mchi^{\rm edge}, \sigmae^{\rm edge})$ (blue markers of Fig.~\ref{fig:exclusion}) and the point of best sensitivity $(\mchi^{\rm best},\sigmae^{\rm best})$ (red markers).}
\label{tab:exclusion}
\begin{ruledtabular}
\begin{tabular*}{\columnwidth}{@{\extracolsep{\fill}} c c cccc}
 & &
 \multicolumn{2}{c}{Low-mass edge} &
 \multicolumn{2}{c}{Best sensitivity} \\
\cline{3-4}\cline{5-6}
$T$ & $E_{\mathrm{th}}$ &
 $\mchi$ & $\sigmae$ &
 $\mchi$ & $\sigmae$ \\

$(^{\circ}\mathrm{C})$ & (keV) &
 (MeV) & $(\mathrm{cm^{2}})$ &
 (MeV) & $(\mathrm{cm^{2}})$ \\
\hline
\multicolumn{6}{c}{Contact \; [$\Fdm=1$]} \\
\hline
$45$ & $0.61$   & $181$ & $9.97\times10^{-23}$ & $1500$ & $8.74\times10^{-36}$ \\
$50$ & $0.3411$ & $103$ & $2.08\times10^{-25}$ & $800$  & $4.94\times10^{-37}$ \\
$55$ & $0.1883$ & $59$  & $1.00\times10^{-28}$ & $450$  & $2.82\times10^{-38}$ \\
$60$ & $0.1018$ & $33$  & $1.66\times10^{-27}$ & $257$  & $1.96\times10^{-39}$ \\
$65$ & $0.0534$ & $19$  & $1.27\times10^{-30}$ & $162$  & $1.93\times10^{-40}$ \\
$70$ & $0.0268$ & $12$  & $5.85\times10^{-36}$ & $101$  & $2.73\times10^{-41}$ \\
\hline
\multicolumn{6}{c}{Long-range \; [$\Fdm=(\aem\me/q)^{2}$]} \\
\hline
$45$ & $0.61$   & $181$ & $2.57\times10^{-14}$ & $1250$ & $2.95\times10^{-27}$ \\
$50$ & $0.3411$ & $103$ & $5.52\times10^{-18}$ & $700$  & $1.67\times10^{-29}$ \\
$55$ & $0.1883$ & $59$  & $2.70\times10^{-22}$ & $400$  & $1.14\times10^{-31}$ \\
$60$ & $0.1018$ & $33$  & $4.78\times10^{-22}$ & $243$  & $1.11\times10^{-33}$ \\
$65$ & $0.0534$ & $19$  & $3.93\times10^{-26}$ & $140$  & $1.43\times10^{-35}$ \\
$70$ & $0.0268$ & $12$  & $2.10\times10^{-32}$ & $83$   & $3.03\times10^{-37}$ \\
\end{tabular*}
\end{ruledtabular}
\end{table}

\subsection{Comparison with the direct-detection landscape}\label{sec:res-compare}
\begin{figure*}[!t]
\centering
  \begin{minipage}{0.81\textwidth}
    \centering
    \includegraphics[width=\linewidth]{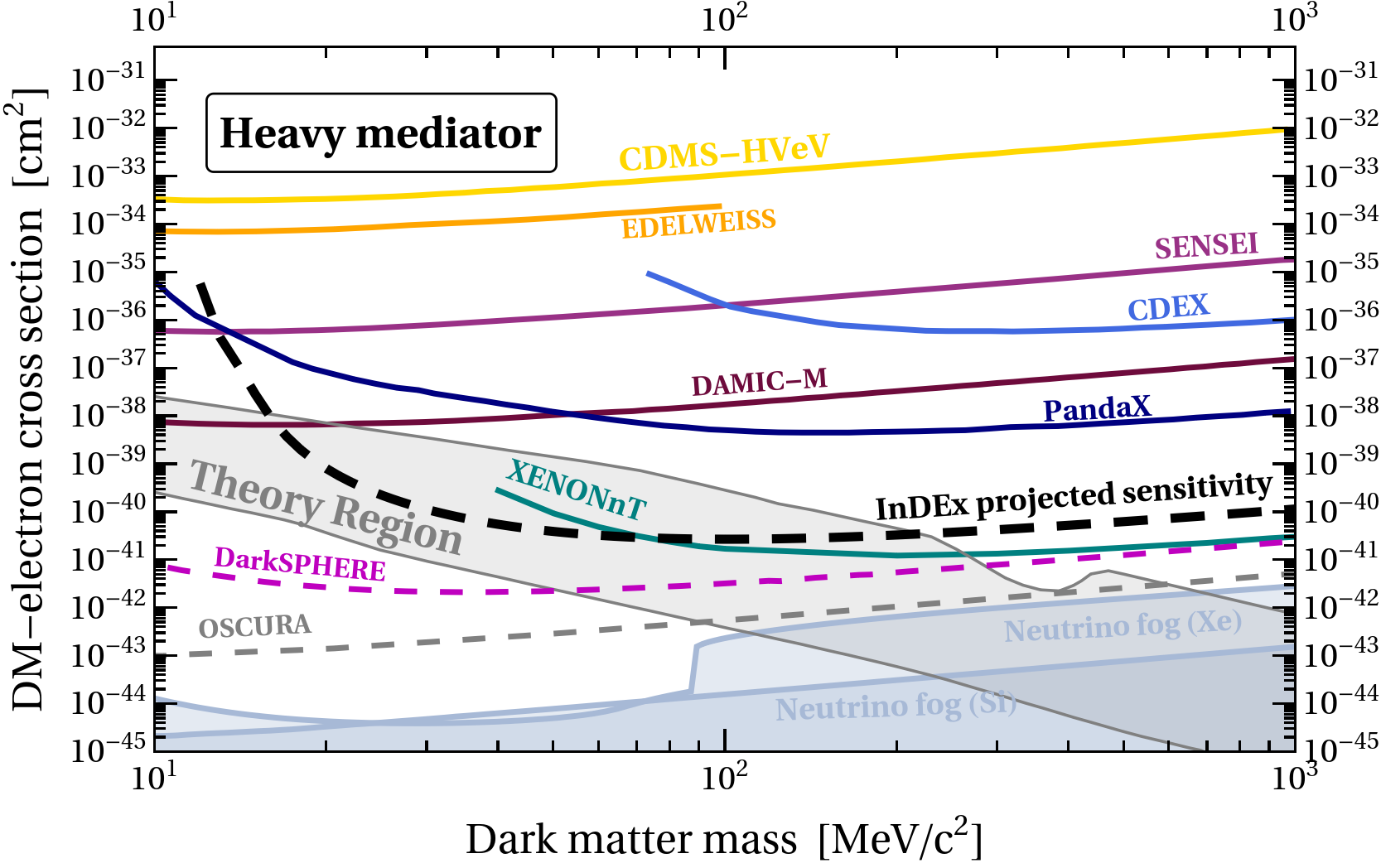}\\[2pt]
    {\small (a) }
  \end{minipage}  
  \begin{minipage}{0.81\textwidth}
    \centering
    \includegraphics[width=\linewidth]{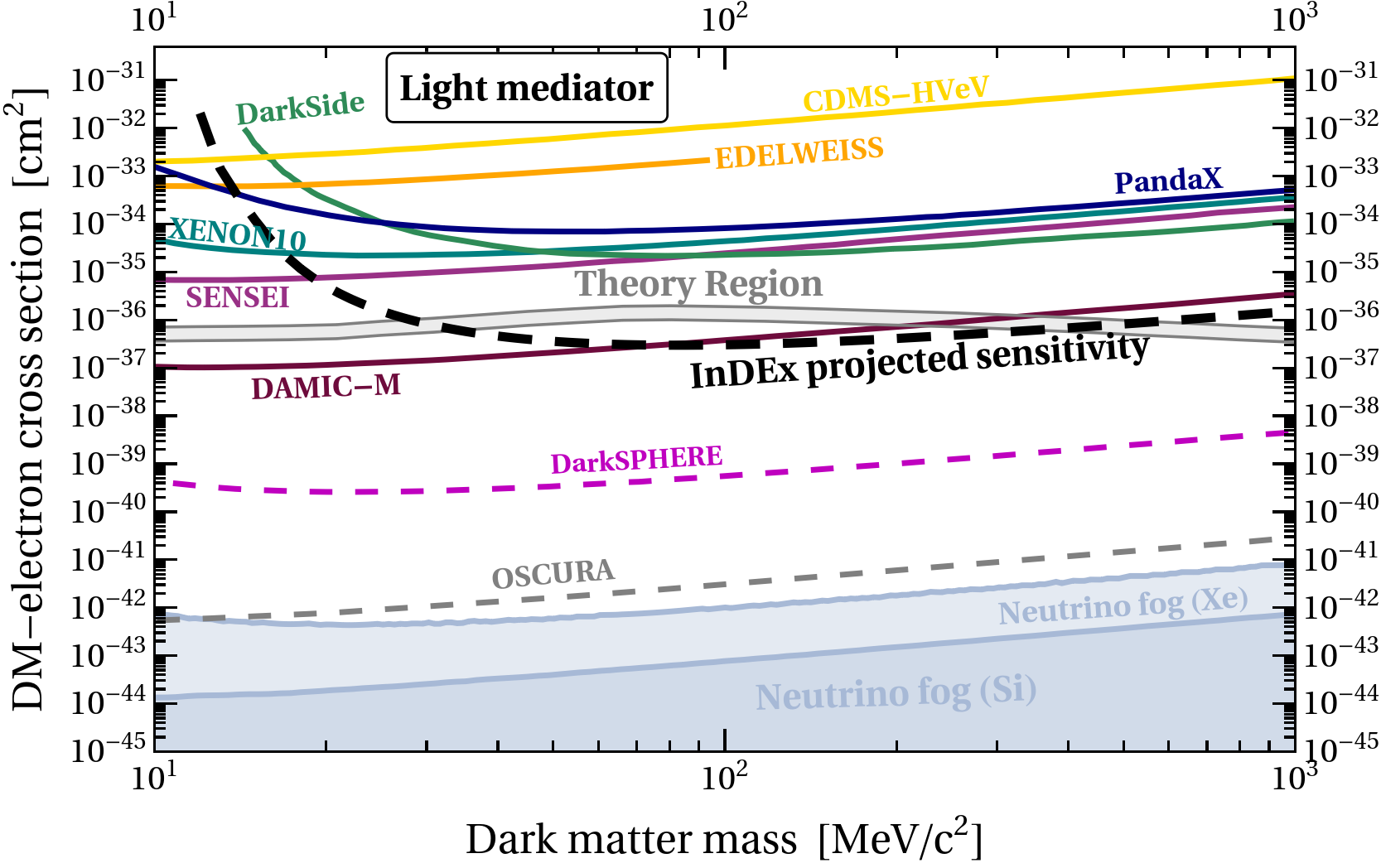}\\[2pt]
    {\small (b) }
  \end{minipage}
\caption{Projected \CHF{} DM--electron exclusion sensitivity (thick dashed black, ``InDEx projected sensitivity'') for the benchmark exposure of $10^{3}~\mathrm{kg\,day}$ at the lowest threshold considered ($E_{\mathrm{th}}=26.8~\mathrm{eV}$, $T=70~^{\circ}\mathrm{C}$), compared with the current landscape of published $90\%$~C.L.\ DM--electron limits and selected projections. Top: heavy-mediator (contact) case, $\Fdm=1$. Bottom: long-range case, $\Fdm=(\aem\me/q)^{2}$. Solid curves correspond to the regions that are excluded, as available for each mediator case, by
XENONnT~\cite{XENONCollaborationP:2026ioh}, XENON10~\cite{essig2012xenon10},
PandaX~\cite{PandaX-II:2021nsg,PandaX:2022xqx,PandaX:2025rrz},
DarkSide-50~\cite{DarkSide:2018bpj,DarkSide:2022knj},
SENSEI~\cite{SENSEI:2020dpa}, DAMIC-M~\cite{DAMIC-M:2023gxo,DAMIC-M:2025luv},
SuperCDMS-HVeV~\cite{SuperCDMS:2018mne,SuperCDMS:2024yiv},
CDEX-10~\cite{CDEX:2022kcd} and EDELWEISS~\cite{EDELWEISS:2020fxc}; dashed gray and magenta curves show the projected reach of Oscura~\cite{Oscura:2022vmi} and
DarkSPHERE~\cite{NEWS-G:2023qwh}. The gray band labeled ``Theory
Region'' indicates the freeze-in relic benchmark~\cite{Hall:2009bx,essig2012, Battaglieri:2017aum}, and the shaded blue regions mark the neutrino fog for xenon and silicon targets. Published limits, projections, theory targets, and neutrino-fog contours were digitized from the compilation of Ref.~\cite{OHare:DirectDetectionPlots}.}
\label{fig:comparison}
\end{figure*}

We now assess the competitiveness of the \CHF{} target against the existing DM--electron landscape, adopting the lowest threshold considered ($E_{\mathrm{th}}=26.8~\mathrm{eV}$ at $T=70~^{\circ}$C) and the benchmark exposure of $10^{3}~\mathrm{kg\,day}$. Figure~\ref{fig:comparison} compares the resulting projections with current constraints. In the heavy-mediator case (top panel), the projected sensitivity extends down to $\mchi\simeq12~\mathrm{MeV}$ (where the limit is $\sigmae=5.85\times10^{-36}~\mathrm{cm}^{2}$), competitive with the lowest masses probed by the latest xenon-ionization searches from XENONnT~\cite{XENONCollaborationP:2026ioh}. Our projection is strongest at $\mchi\simeq101~\mathrm{MeV}$, where it reaches $\sigmae=2.73\times10^{-41}~\mathrm{cm}^{2}$ and dips into the freeze-in relic ``theory region''. For the masses between $\mchi\simeq20~\mathrm{MeV}$ and $\mchi\simeq80~\mathrm{MeV}$ the \InDEx{}
curve lies below \emph{all} constraints shown -- including the strongest silicon skipper-CCD and xenon bounds of DAMIC-M~\cite{DAMIC-M:2023gxo}, PandaX~\cite{PandaX-II:2021nsg,PandaX:2022xqx} and the latest XENONnT~\cite{XENONCollaborationP:2026ioh}. In the long-range case (bottom panel), the projection reaches $\mchi=12~\mathrm{MeV}$ (at $\sigmae=2.10\times10^{-32}~\mathrm{cm}^{2}$) with best sensitivity $\sigmae=3.03\times10^{-37}~\mathrm{cm}^{2}$ at $\mchi\simeq83~\mathrm{MeV}$, comparable to the leading long-range limit of DAMIC-M~\cite{DAMIC-M:2023gxo,DAMIC-M:2025luv} over much of the sub-GeV window. We note that the most recent xenon and skipper-CCD analyses~\cite{PandaX:2025rrz,DAMIC-M:2025luv} continue to strengthen the region $\mchi\gtrsim100~\mathrm{MeV}$ -- PandaX-4T, for instance, now reports $\sigmae\simeq1.5\times10^{-41}~\mathrm{cm^{2}}$ at $200~\mathrm{MeV}$ for the contact case~\cite{PandaX:2025rrz}. The present projection is best regarded as competitive with the current frontier at those masses. Below a few tens of MeV, the low-threshold superheated liquid retains a distinctive reach. Lower thresholds, reachable by operating at still higher temperatures, would move the curves further down and to the left, at the cost of the background growth discussed in Sec.~\ref{sec:detector}.

Two further points are worth emphasizing. First, these are best-case projections, obtained under the zero-background, unit-efficiency assumptions of Secs.~\ref{sec:detector} and \ref{sec:limits}; a realistic assessment must fold in the measured electron-recoil nucleation response, background, and efficiency of the \InDEx{} detectors in the gamma-sensitive regime. Even with this caveat, the projection demonstrates that a superheated refrigerant target -- chemically and operationally distinct from the semiconductor and noble-liquid detectors that dominate the field, and already deployed underground -- can, with achievable thresholds and a moderate exposure of $10^{3}~\mathrm{kg\,day}$, reach the cross-section scales relevant for sub-GeV DM. Second, within \InDEx{} itself, the electron channel is complementary to the nuclear-recoil program of Refs.~\cite{seth2020,Kumar:2025ofs,das2025index} at the level of the \emph{couplings} probed. Nuclear recoils are essentially blind to DM that couples predominantly to leptons, so the present analysis extends the class of DM models testable with the same detector, rather than merely the accessible mass range.

\section{Summary and outlook}\label{sec:summary}

We have evaluated, for the first time, the sensitivity of the superheated \CHF{} target of \InDEx{} to dark matter--electron scattering. The electron channel becomes available once the detectors are operated above the measured onset of electron-recoil--induced bubble nucleation at $(38.5\pm1.4)~^{\circ}$C~\cite{sahoo2019gamma}; we defined a six-point operating window, $T=45$--$70~^{\circ}$C, whose lower edge lies safely above this onset and whose upper edge balances the sub-GeV mass reach of the $26.8~\mathrm{eV}$ Seitz threshold against the growth of the gamma-induced background with temperature (Sec.~\ref{sec:detector}). The ionization response of the target was computed in the isolated-atom approximation with Roothaan--Hartree--Fock bound states~\cite{Bunge:1993jsz} and hydrogenic continuum states, evaluated with the \texttt{DarkART} code~\cite{DarkART} and validated against the published argon and xenon responses of Ref.~\cite{catena2020}. For a dark-photon-type mediator and a benchmark exposure of $10^{3}~\mathrm{kg\,day}$, the projected $90\%$~C.L.\ limits at the deepest threshold extend down to $\mchi\simeq12~\mathrm{MeV}$ and reach best sensitivities of $\sigmae\simeq2.73\times10^{-41}~\mathrm{cm^{2}}$ (contact, near $101~\mathrm{MeV}$) and $\sigmae\simeq3.03\times10^{-37}~\mathrm{cm^{2}}$ (long-range, near $83~\mathrm{MeV}$), competitive with the leading electron-scattering experiments and complementary to the \InDEx{} nuclear-recoil program in both mass reach and couplings probed.

These projections rest on three working assumptions, stated explicitly in Secs.~\ref{sec:detector} and \ref{sec:signal}: (i)~the electronic structure of the target is approximated by a stoichiometry-weighted superposition of isolated C, H, and F atoms, neglecting $\mathcal{O}(10~\mathrm{eV})$ bonding effects on the valence orbitals; (ii)~the Seitz critical energy is used as an effective deposited-energy threshold for electron recoils, with unit efficiency above threshold, although electron-recoil nucleation is ionization-driven and may deviate from the Seitz scaling~\cite{pico2019electron}; and (iii)~zero background is assumed over the full exposure. Correspondingly, the immediate next steps of this study are following: A dedicated quantum-chemical treatment of the bound \CHF{} system, along the lines now available for organic targets~\cite{Blanco:2019lrf}, together with an extension of the analysis to general effective operators beyond the dark-photon benchmark~\cite{catena2020,catena2023beyond,liang2024}. With these elements in place, the electron channel developed here can be applied directly to forthcoming \InDEx{} data, extending the physics reach of the experiment to sub-GeV dark matter and to leptophilic couplings with the detectors already in hand.

\begin{acknowledgments}
The authors acknowledge DAE, Saha Institute of Nuclear Physics (SINP), Kolkata, as their research funding source. The authors thank Ms. Kaynat Fatima for providing useful information, and the members of the \InDEx{} collaboration for helpful discussions.
\end{acknowledgments}

\appendix

\section{Electronic-structure calculation of \CHF}\label{app:estructure}
As described in Sec.~\ref{sec:ff}, the electronic structure of \CHF{} is modeled as an isolated-atom superposition. For carbon and
fluorine, the radial part of each occupied orbital is the Roothaan--Hartree--Fock (RHF) ground-state solution of Ref.~\cite{Bunge:1993jsz}, expanded on Slater-type orbitals,
\begin{equation}
\begin{split}
  R_{n\ell}(r)\;&=\;\sum_{j} C_{j\ell n}\,{\cal N}_{j\ell}
  \left(\frac{r}{a_{0}}\right)^{\!n'_{j\ell}-1}
  e^{-Z_{j\ell}\,r/a_{0}}\,,\\
  {\cal N}_{j\ell}\;&=\;a_{0}^{-3/2}\,
  \frac{\bigl(2Z_{j\ell}\bigr)^{\,n'_{j\ell}+1/2}}
       {\sqrt{\bigl(2n'_{j\ell}\bigr)!}}\,,
  \label{eq:STO}
\end{split}
\end{equation}
where $a_{0}$ is the Bohr radius and the coefficients $C_{j\ell n}$,
$Z_{j\ell}$ and $n'_{j\ell}$ are tabulated in Ref.~\cite{Bunge:1993jsz}
[cf.\ Eq.~(B33) of Ref.~\cite{catena2020}]; the hydrogen $1s$ orbital is
exact. Binding energies are identified with the RHF orbital eigenvalues
(Koopmans' theorem). The final-state electron of asymptotic momentum $k'$
and angular momentum $\ell'$ is the positive-energy solution of the radial
Schr\"odinger equation in the Coulomb potential of the residual ion with
effective charge $Z_{\mathrm{eff}}^{n\ell}=n\,\sqrt{E_{B}^{n\ell}/13.6~\mathrm{eV}}$,
normalized as in Appendix~B4 of Ref.~\cite{catena2020}. Table~\ref{tab:orbitals} lists the orbitals included in the analysis together with their binding energies, occupancies, and effective charges.
\begin{table}[t]
  \centering
  \caption{Atomic orbitals of the \CHF{} constituents included in
  the analysis: binding energies $E_{B}^{n\ell}$ (RHF eigenvalues of
  Ref.~\cite{Bunge:1993jsz}; hydrogen exact), occupancy per
  atom, number of atoms per \CHF{} unit $\nu_{A}$, and effective charge
  $Z_{\mathrm{eff}}^{n\ell}=n\sqrt{E_{B}^{n\ell}/13.6~\mathrm{eV}}$ of the
  final-state Coulomb wavefunctions. Each \CHF{} unit contains 50
  electrons in total.}
  \label{tab:orbitals}
  \begin{tabular}{l c c c c}
    \hline\hline
    Orbital & $E_{B}^{n\ell}$ [eV] & Occupancy & $\nu_{A}$ & $Z_{\mathrm{eff}}^{n\ell}$ \\
    \hline
    H  $1s$ & 13.61  & 1 & 2 & 1.00 \\
    C  $1s$ & 308.18 & 2 & 2 & 4.76 \\
    C  $2s$ & 19.20  & 2 & 2 & 2.38 \\
    C  $2p$ & 11.79  & 2 & 2 & 1.86 \\
    F  $1s$ & 717.92 & 2 & 4 & 7.26 \\
    F  $2s$ & 42.79  & 2 & 4 & 3.55 \\
    F  $2p$ & 19.87  & 5 & 4 & 2.42 \\
    \hline\hline
  \end{tabular}
\end{table}
\bibliographystyle{apsrev4-2}
\bibliography{refs}

\end{document}